\documentclass[journal]{IEEEtran}
\AtBeginDocument{%
  \setlength\abovedisplayskip{2.5pt}
  \setlength\belowdisplayskip{2.5pt}}

\usepackage[T1]{fontenc}
\usepackage[utf8]{inputenc}
\usepackage{graphicx}
\usepackage{hyperref}
\usepackage{multirow}
\usepackage{tabularx}
\usepackage{color}
\usepackage{textcomp}
\usepackage{tipa}
\usepackage{amsmath}
\usepackage{amssymb}
\usepackage{amsfonts}
\usepackage{amsxtra}
\usepackage{isomath}
\usepackage{mathtools}
\usepackage{txfonts}
\usepackage{upgreek}
\usepackage{enumerate}
\usepackage{tensor}
\usepackage{pifont}
\usepackage{soul}
\usepackage{arydshln}
\usepackage{cite}
\usepackage{algorithmic}
\usepackage{bbding}
\usepackage{booktabs}
\usepackage[squaren]{SIunits}
\usepackage{longtable}
\usepackage{rotating}
\usepackage{makecell}
\usepackage{pdflscape}
\usepackage{adjustbox}
\usepackage{bm,array}
\usepackage[dvipsnames]{xcolor}
\usepackage{caption}
\usepackage{color,soul}
\usepackage[roman]{parnotes}
\usepackage{balance}
\begin{document}

\title{Reradiation and Scattering from a Reconfigurable Intelligent Surface: A General Macroscopic Model}

\renewcommand\footnotemark{}
\renewcommand\footnoterule{}

\author{
Vittorio~Degli-Esposti,~\IEEEmembership{Senior~Member,~IEEE},
Enrico~M.~Vitucci,~\IEEEmembership{Senior~Member,~IEEE},
Marco~Di~Renzo,~\IEEEmembership{Fellow,~IEEE},
and~Sergei~Tretyakov,~\IEEEmembership{Fellow,~IEEE}
\vspace{-0.75cm} }
\thanks{V. Degli-Esposti and E. M. Vitucci are with the Dept. of Electrical, Electronic and Information Engineering "G. Marconi", CNIT, University of Bologna, 40136 Bologna, Italy (e-mail: \{v.degliesposti,enricomaria.vitucci\}@unibo.it).}
\thanks{M. Di Renzo is with Universit\'{e} Paris-Saclay, CNRS, CentraleSup\'{e}lec, Laboratoire des Signaux et Syst\`{e}mes, 91192 Gif-sur-Yvette, France (e-mail: marco.di-renzo@universite-paris-saclay.fr). The work of M. Di Renzo was supported in part by the European Commission through the H2020 ARIADNE project under grant agreement number 871464 and through the H2020 RISE-6G project under grant agreement number 101017011.}
\thanks{S. Tretyakov is with the Department of Electronics and Nanoengineering, School of Electrical Engineering, Aalto University, 02150 Espoo, Finland (e-mail: sergei.tretyakov@aalto.fi).}

\maketitle

\begin{abstract}
Reconfigurable Intelligent Surfaces (RISs) have attracted major attention in the last few years, thanks to their useful characteristics. An RIS is a nearly passive thin surface that can dynamically change the reradiated field, and can therefore realize anomalous reflection, refraction, focalization, or other wave transformations for engineering the radio propagation environment or realizing novel surface-type antennas. Evaluating the performance and optimizing the deployment of RISs in wireless networks need physically consistent frameworks that account for the electromagnetic characteristics of dynamic metasurfaces. In this paper, we introduce a general macroscopic model for evaluating the scattering from an RIS. The proposed method decomposes the wave reradiated from an RIS into multiple scattering contributions and is aimed at being embedded into ray-based models. Since state-of-the-art ray-based models can already efficiently simulate specular wave
reflection, diffraction, and diffuse scattering, but not anomalous reradiation, we enhance them with an approach based on Huygens' principle and propose two possible implementations for it.
Multiple reradiation modes can be modeled through the proposed approach, using the power conservation principle. We validate the accuracy of the proposed model by benchmarking it against several case studies available in the literature, which are based on analytical models, full-wave simulations, and measurements.
\end{abstract}


\maketitle

\section{Introduction}
\label{sec1}
With the current deployment of fifth generation (5G) communication systems, it is now a critical time to identify enabling technologies for the sixth generation (6G). 6G systems are expected to fulfill more stringent requirements than 5G networks in terms of transmission capacity, reliability, latency, coverage, energy consumption, and connection density. Existing 5G technologies, such as millimeter-wave communications, massive multiple-input multiple-output schemes, and ultra-dense heterogeneous networks, are mainly focused on system designs at the transmitter and receiver sides, as well as on the deployment of additional network infrastructure elements with power amplification, digital signal processing capabilities, and backhaul availability.
The purpose of currently available 5G technologies is mainly to capitalize on or to cope with often unfavorable wireless propagation environments. In fact, the propagation environment is conventionally modeled as an exogenous entity that cannot be controlled but can only be adapted to. According to this design paradigm, communication engineers usually design the transmitters, receivers, and transmission protocols based on the specific properties of the wireless channels and for achieving the desired performance.

Recently, the technology referred to as reconfigurable intelligent surface (RIS) has emerged as a promising option for its capability of customizing the wireless propagation environment through nearly passive signal transformations. An RIS is a thin surface that is engineered to possess properties that enable it to dynamically control the electromagnetic waves through, e.g., signal reflections, refractions, focusing, and their combinations.

In wireless communications, RISs are intended to realize so-called programmable and reconfigurable wireless propagation environments, i.e., large-scale or small-scale propagation environments that are not viewed and treated as random uncontrollable entities but become part of the network design parameters that are subject to optimization for fulfilling the stringent requirements of 6G networks \cite{ref1,ref2,ref3,ref4}. Recent applications of RISs in wireless communications include their use as nearly passive relay-type surfaces, multi-stream multi-antenna transmitters, and reconfigurable ambient backscatters that work without requiring power amplification or digital signal processing \cite{ref5,ref6,ref7}.

An RIS operates, on the other hand, in the electromagnetic domain directly on the electromagnetic waves that impinge upon it. The performance evaluation and optimization deployment of RISs in wireless networks require physically consistent and realistic models that account for their electromagnetic characteristics and physical implementations, which include the wave transformations that they realize, the size, losses, parasitic effects, and transmission distances \cite{ref5}. Accurate microscopic simulations with the aid of full-wave electromagnetic models  may be utilized as well. These latter models and methods are, however, too demanding in terms of computational resources and this may prevent their utilization for link- or system-level simulations in wireless networks \cite{Huawei}.

Motivated by these considerations, a few research works have recently investigated macroscopic methods for modeling the scattering from finite-size RISs, which are based on different analytical approaches and assumptions.
A summary of the available contributions and a brief description of their main features and limitations are available in Table~\ref{Tab:1} \cite{ref8,ref9,ref10,ref11,ref12,ref13,ref14,ref15,ref16,ref17,ref18,ref19}.

A more extensive state-of-the art review can be found in \cite{ref18,ref19}.

\begin{table*}[!h]
\caption{Summary of state-of-the-art contributions with their main features and limitations}
\label{Tab:1}
\begin{tabularx}{\textwidth}{|
p{\dimexpr 0.09\linewidth-2\tabcolsep-2\arrayrulewidth}|
p{\dimexpr 0.59\linewidth-2\tabcolsep-\arrayrulewidth}|p{\dimexpr 0.32\linewidth-2\tabcolsep-\arrayrulewidth}|} \hline
\centering\textbf{\textit{Reference}} & \centering\textbf{\textit{Main features}} & \centering\arraybackslash{}\textbf{\textit{Limitations}} \\\hline
\cite{ref8} & \raggedright\arraybackslash{}- Physical optics and the scalar Huygens-Fresnel principle are used for analysis \par - Asymptotic scaling laws as a function of the distance and surface size are derived \par - Far-field and near-field case studies are considered and discussed & - A two-dimensional space is considered \par - Parasitic modes and diffuse scattering are not considered\\\hline
\cite{ref9}, \cite{ref10} & \raggedright\arraybackslash{}- Antenna theory and a locally periodic model for the surface are used for analysis \par - The model is validated with measurements using manufactured RISs & - Parasitic modes and diffuse scattering are not considered \\\hline
\cite{ref11} & \raggedright\arraybackslash{}- Antenna theory is used for analysis \par - Far-field and near-field case studies are considered and discussed & - Only reflectarray-type RISs are studied \par - Only the far-field case is considered\\\hline
\cite{ref12} & \raggedright\arraybackslash{}- Physical optics and antenna theory are used for analysis in the far-field region \par - Path-loss behavior of anomalous reflectors is analyzed and discussed & - Only the surface electric currents are modeled \par - Only anomalous reflection is considered\\\hline
\cite{ref13} & \raggedright\arraybackslash{}- Methods for electronically steerable parasitic array radiators are used for analysis \par - Scaling laws as a function of the transmission distance are discussed & - The analysis is limited to large antenna-arrays acting as mirrors or scatterers\\\hline
\cite{ref14} & \raggedright\arraybackslash{}- Physical optics methods are used for analysis \par - The RIS is modeled as a multi-tile surface made of perfectly magnetic conductors \par - A scalable optimization algorithm is proposed for a multi-tile anomalous reflector & - Only anomalous reflection is considered \par - Parasitic effects and diffuse scattering are not considered\\\hline
\cite{ref15},  \par \cite{ref16}, \cite{ref17} & \raggedright\arraybackslash{}- A model based on impedance-controlled thin dipole antennas is proposed \par - The mutual coupling and the impact of the tuning elements are considered \par - Optimization algorithms are introduced to exploit the mutual coupling & \par - Minimum scattering antenna elements are considered
\par - Parasitic modes and diffuse scattering are not considered\\\hline
\cite{ref18} & \raggedright\arraybackslash{}- Physical optics and the vector Huygens-Fresnel principle are used for analysis \par - Asymptotic scaling laws as a function of the distance and surface size are derived \par - Far-field and near-field case studies are considered and discussed \par - Multiple RIS functions are considered, including anomalous reflection and focusing \par - The surface can operate in reflection and refraction mode & - Diffuse scattering is not considered \\\hline
\cite{ref19} & \raggedright\arraybackslash{}- Physical optics and the vector Huygens-Fresnel principle are used for analysis \par - A non-ideal multi-mode scattering model based on Floquet’s theory is proposed \par - The model is validated against full-wave simulations & - Applicable only to periodic surfaces \par - Diffuse scattering is not considered \\\hline
This work & \raggedright\arraybackslash{}- A parametric scattering model based on a power balance conservation principle is proposed \par - The scattering model accounts for specular reflection, multi-mode anomalous reradiation, and diffuse scattering; extension to transmission and curved RISs is possible \par - Two reradiation models based on the vector Huygens-Fresnel principle and antenna theory are analyzed and compared \par - The integration of the proposed scattering model with ray tracing is discussed \par - The model is validated against Floquet’s theory, simulations, and empirical data & - The model needs to be parameterized through measurements or full-wave simulations \\\hline
\end{tabularx}
\end{table*}

\begin{figure*}[!ht]
\centering
\includegraphics[width=0.7\textwidth]{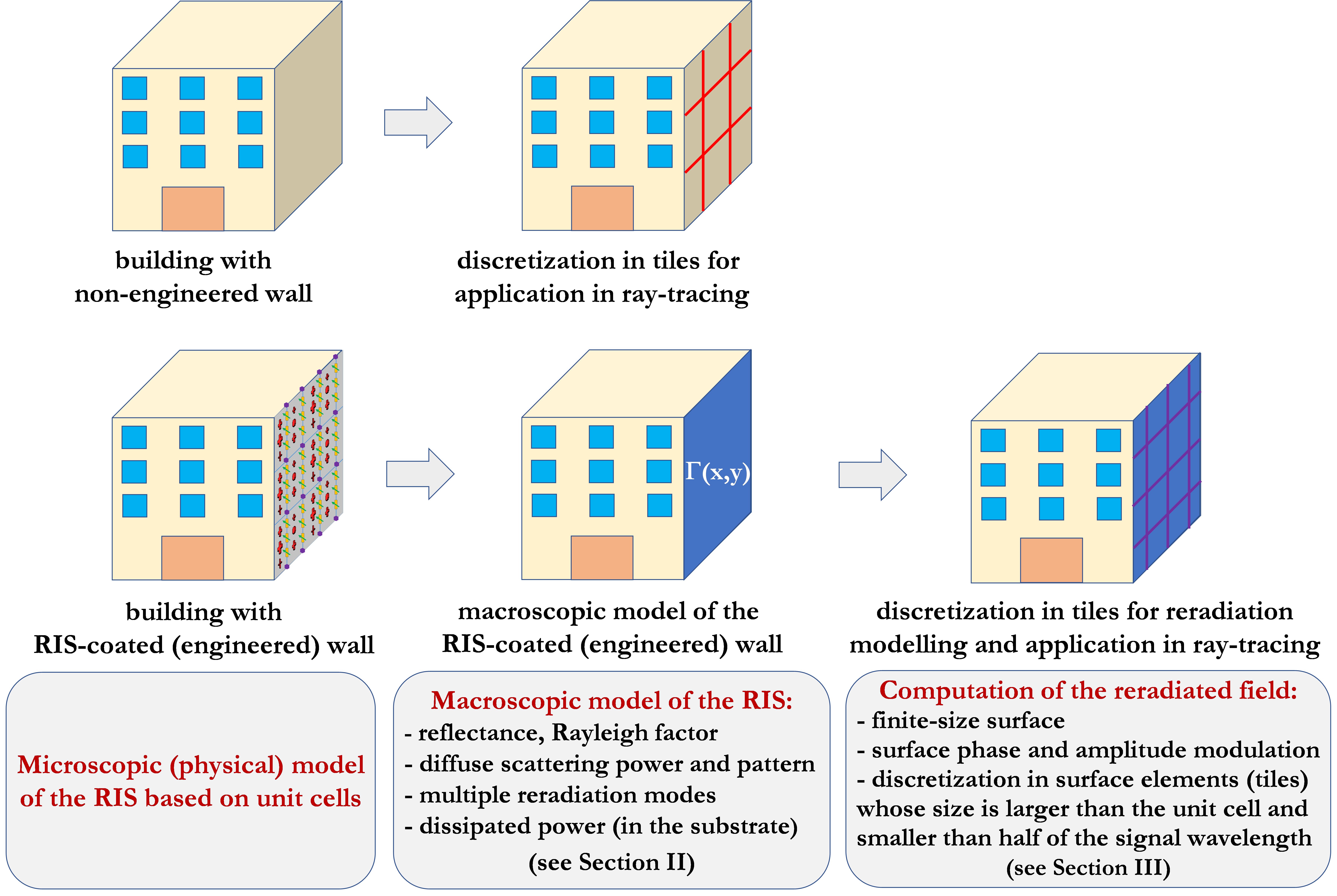}
\caption{Comparison between non-engineered walls and RIS-coated walls: Proposed methodology and integration in ray-based models.}
\label{fig:1} \vspace{-0.25cm}
\end{figure*}

Based on Table~\ref{Tab:1}, we evince that, except for \cite{ref18} and \cite{ref19}, the contributions available to date on modeling the scattered electromagnetic field from finite-size RISs can be referred to as \textit{ideal scattering models}. \textcolor{black}{More precisely, the term ``ideal'' refers to the assumptions that (i) the RIS reflects the incident radio waves towards a single specified direction (dominant) without generating parasitic scattered waves towards (unwanted) directions, and (ii) the RIS is illuminated by a single wave that impinges from a single direction of design.} These two major assumptions can be removed, under some assumptions, by considering the analytical models proposed in \cite{ref18} and \cite{ref19}. In \cite{ref18}, Theorem 2 can be applied to any field $E_{S}$ on the surface of the RIS, which can be even obtained from electromagnetic simulations and can account for multiple scattering modes. However, the subsequent analysis applied to reflective and refractive RISs is performed under the assumption that only a single (the dominant) scattering mode exists. The analysis reported in \cite{ref19} explicitly accounts for the existence of multiple directions of scattering (or reradiation) based on Floquet’s theory. This latter theory is, however, applicable only to periodic metasurfaces. Also, the analysis is specialized to the far-field region of the RIS. Neither in \cite{ref18} nor in \cite{ref19}, the authors consider the presence and impact of diffuse scattering that may be caused by, e.g., design tradeoffs, construction inaccuracies and the deposit of dust. Furthermore, the scattering models proposed for RISs to date are based either on Huygens' principle under the physical optics approximation regime (e.g., \cite{ref18}, \cite{ref19}) or on antenna theory (e.g., \cite{ref9}, \cite{ref10}). However, it is unclear to what extent these two methods can be applied and provide similar results. Therefore, we evince that understanding and realistically modeling the scattering from finite-size RISs that can apply general wave transformations, realized through periodic or aperiodic surfaces, are open research problems.

In contrast to the current state-of-the-art, in the present paper we introduce an approach for modeling the scattering from a general finite-size and non-ideal RIS. The model is conceived for being integrated into currently available ray-based models, such as ray tracing and ray launching methods, which are recognized as the most suitable and efficient deterministic models for realistic radio propagation simulations in man-made environments.
Several ray-based models discretize ordinary surfaces into surface elements (also called ``tiles'') in order to simulate diffuse scattering (by using, e.g., the effective roughness (ER) model \cite{ref20}) and/or to achieve a good computational efficiency through parallel computing algorithms \cite{ref21}. In the present work, we leverage the tile-based approach to simulate the anomalous scattering from an RIS (also referred to as anomalous reflection or reradiation in the sequel), by using a method based on Huygens' principle \cite{ref22,ref23}.
The proposed methodology for modeling an RIS within a ray-based propagation simulator is sketched in Fig.~\ref{fig:1}.  While ordinary surfaces are simply discretized to apply efficient ray-based models, RIS-coated surfaces are first homogenized and described through a proper spatial modulation function $\Upgamma(x,y)$, which accounts for anomalous reradiation, and are then discretized to apply computation procedures similar to those utilized for ordinary surfaces. The scattering from an RIS surface is therefore decomposed into ``typical'' scattering effects, such as specular reflection, diffraction, diffuse scattering, and anomalous reradiation. While the former effects are treated by using well-established theories and methods, such as geometrical optics, the uniform theory of diffraction, and the ER model, the anomalous reradiation is treated by using Huygens' principle approach, by integrating it into currently available frameworks for the discretization of surfaces and for efficient computation. This approach is described in Section~\ref{sec3}. The key feature of the proposed model consists of fulfilling the power balance between the different scattering modes, which is ensured by using a parameter-based approach. Specifically, the model is based on two steps: i) the definition of the global power balance between conventional and anomalous scattering modes, and ii) the computation of the scattered field as a coherent sum of multiple contributions, including conventional and anomalous reradiated modes.
In more detail, the main contributions of this paper are as follows:
\begin{itemize}
\item
We introduce a general parametric approach for modeling the scattering from a finite-size RIS, which is suitable for integration into ray-based models.

\item
The proposed model is macroscopic, as it is agnostic to the specific microscopic (unit cells) physical implementation of the RIS, and is instead characterized by macroscopic parameters.

\item
The model explicitly takes into account diffuse scattering, in addition to the desired and undesired reradiation modes, it is not limited to flat metasurfaces, and it can be easily generalized for application to refractive metasurfaces (not considered here for brevity).

\item
We consider and compare two versions of the model for computing the anomalous reradiated field.

\item
The method is conceived to be integrated into advanced discrete ray-based models \cite{ref21}, and it can be efficiently implemented on parallel computing platforms.

\item
The model is tested and validated against results available in the literature, which are based on theory, full-wave simulations, and measurements conducted on manufactured RISs. The results confirm the generality and accuracy of the proposed approach, as well as the non-negligible impact that multi-mode reradiation and diffuse scattering may have on the total scattered field and, notably, on the far-field radiation pattern of a finite-size RIS.

\end{itemize}

The rest of this paper is organized as follows. In Section~\ref{sec2}, we describe the proposed macroscopic scattering model and the power balance requirement among the different scattering modes. In Section~\ref{sec3}, we introduce and compare two versions of the proposed method for computing the reradiated field from a finite-size RIS. In Section~\ref{sec4}, we validate the proposed macroscopic scattering model with the aid of numerical simulations. Finally, Section~\ref{sec5} concludes the paper.

\section{A Macroscopic Scattering Model for RISs}
\label{sec2}

Even disregarding the impact of the finite size of the surface and the near-field illumination, real-world metasurfaces can generate multiple reradiated modes but only one of them is usually the desired mode. For example, let us consider an anomalous reflector that is designed, under a plane wave illumination, to reradiate a single plane wave towards a non-specular direction. It is known that a non-local design is usually needed for realizing  surfaces with a high reradiation efficiency and for avoiding parasitic modes \cite{ref24}. The use of different design methods, such as designing phase-gradient metasurfaces based on the locally periodical design, may result in a low reradiation efficiency and several parasitic modes, as dictated by Floquet's principle \cite{PIEEE}. For example, specular reflection is one of the most relevant parasitic modes. In addition, diffuse scattering effects due to design trade-offs, construction inaccuracies, and the deposit of dust or raindrops on the surface may affect the reradiation efficiency as well.

Currently available ray-based simulation tools are conceived for modeling specular reflection, diffraction and diffuse scattering \cite{ref25}. On the other hand, they cannot model anomalous reradiation. The proposed scattering model for RISs constitutes a plug-in extension of ray-based models, which accounts for several physical implementations of RISs and wave transformations that they can apply. In the proposed model, an RIS is partitioned into surface elements (or ``tiles''),  which is shown in Section~\ref{sec3} to be a suitable approach for efficiently computing the reradiated field. In general, however, the surface element does not correspond to either a single meta-atom or unit cell of the metasurface or to any other physical element of the microscopic implementation of the RIS.

The proposed approach for modeling the scattering from an RIS starts from imposing a power balance constraint, according to the power conservation principle, between the incident and the reradiated waves. The power balance constraint is imposed regardless of the size of the RIS, and is formulated in terms of macroscopic parameters that measure the relative intensity of all possible reradiated and scattered modes, which carry power towards different directions. As far as periodic surfaces are concerned, the directions and the amplitudes of the different reradiated modes can be obtained from Floquet’s theory and by applying the mode-matching approach \cite{ref19}. As far as aperiodic surfaces are concerned, the reradiated modes and their corresponding amplitudes and phases are usually obtained through full-wave simulations or measurements. For both periodic and aperiodic surfaces, the  reradiated modes depend on the angle of incidence and the specific design of the surface, e.g., the surface impedance.

The proposed power balance criterion accounts for specular reflection, diffraction, diffuse scattering, and for all non-specular reradiated modes. In the proposed model, however, we ignore the field reradiated by evanescent (non-propagating) modes. This implies that the  approach is applicable to observation points that are at least a few wavelengths away from the surface of the RIS, where the impact of the evanescent modes can be safely assumed to be negligible. However, the proposed approach can be applied to both the near-field and far-field regions of the RIS structure.

After ensuring the power conservation principle, the scattered field is computed by taking into account the finite size of the surface and the characteristics of every possible reradiated mode. Specifically, anomalous reradiation is computed by adopting two methods: (1)  Huygens' principle based on a generalization, proposed in this paper,  of the method of image currents \cite{ref22,ref23} (see Section~\ref{subsec3A}), and (2) antenna theory according to which an RIS is modeled as a planar array of antennas (see Section~\ref{subsec3B}). Both methods use as an input the parameters obtained from the power conservation principle.

The proposed approach is inspired to the ER model \cite{ref20}, which originally enforces the power balance principle to the specular reflection and diffuse scattering. Therefore, we first briefly recall the original ER model, and then generalize it to account for multi-mode anomalous reradiation.

\subsection{The Effective Roughness Model}
\label{subsec2B}

\begin{figure}
\centering
\includegraphics[width=0.4\textwidth]{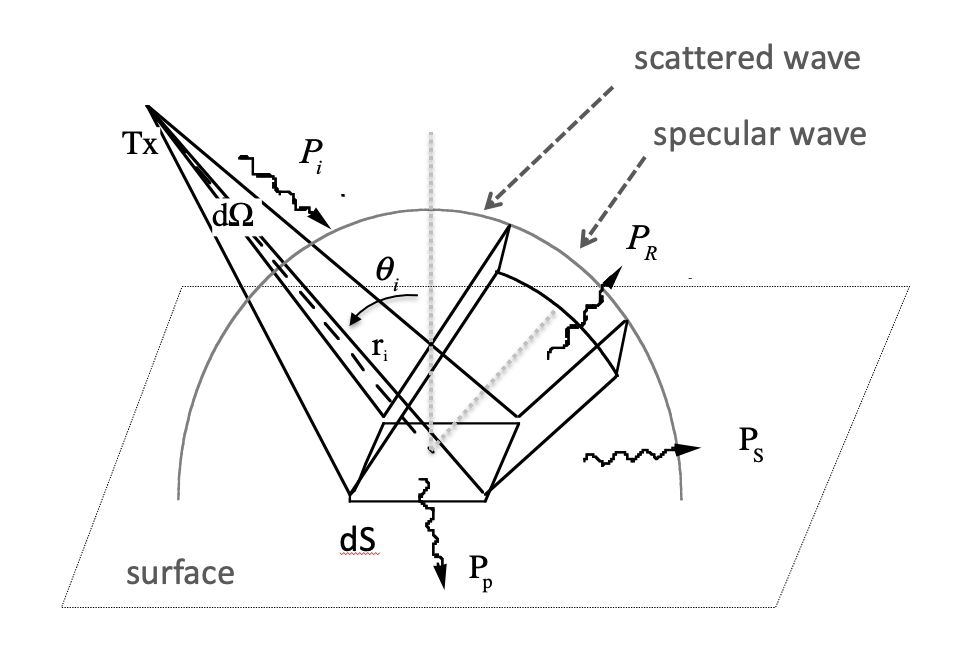}
\caption{Local power balance for the ER model: the incident power \textit{P}$_{i}$ is split into specularly reflected (\textit{P}$_{r}$), diffuse (\textit{P}$_{s}$), and transmitted (\textit{P}$_{p}$) power.}
\label{fig:2} \vspace{-0.25cm}
\end{figure}

The ER model is a heuristic approach for modeling diffuse scattering from ordinary surfaces (e.g., building walls) that can be easily integrated into ray-based field prediction algorithms \cite{ref25}. As illustrated in Fig.~\ref{fig:2}, the ER model is based on a power balance principle applied to a generic surface element (``tile'') of the wall. In Fig.~\ref{fig:2}, the generic tile is denoted by \textit{dS}. A wave that impinges upon the tile is assumed to generate both a specularly reflected wave and a diffuse scattered wave. In addition, some power penetrates into the wall. The field scattered by each tile is modeled as a non-uniform spherical wave that departs from the tile itself and propagates in the same half-space as the incident wave. The intensity of the scattered wave is determined by a scattering coefficient \textit{S} and by a scattering pattern that depends on the irregularities of the wall, as discussed next.

Based on Fig.~\ref{fig:2} and assuming that the surface is illuminated by an electromagnetic wave whose direction of incidence is $\theta _{i}$ and whose electric field is \textbf{E}$_{\mathrm{i}}$, the following power balance law holds:
\begin{equation}
\tag{1}
\begin{split}
P_{i}=\frac{\left| \mathbf{E}_{i}\right| ^{2}}{2\eta }\cos \left(\theta _{i}\right)dS=P_{r}+P_{s}+P_{p}
\end{split}
\end{equation}
where \textit{P}$_{i}$\textit{, P}$_{r}$\textit{, P}$_{s}$\textit{,} are the incident, specularly reflected, and diffuse scattered powers at the generic surface element \textit{dS}, and \textit{P}$_{p}$ is the power that penetrates into the wall. More precisely, the model assumes that the wall has some surface irregularities with respect to an ideal uniform flat surface, that divert a fraction of the power from the direction of specular reflection towards other directions. The dissipated power \textit{P}$_{p}$ is assumed to be independent of these irregularities.

Let us introduce the scattering parameter $S$ (0{\textless}\textit{S}{\textless}1), which is defined so that $S^{2}$ corresponds to the ratio between the diffuse scattering power and the reflected power. Also, let us introduce the specular field-reduction factor \textit{R} (0{\textless}\textit{R}{\textless}1, which is often called Rayleigh factor) to account for the attenuation that the reflected wave undergoes with respect to a smooth flat layer. From (1) we obtain
\begin{equation}
\tag{2}
P_{i}=R^{2}\rho P_{i}+S^{2}\rho P_{i}+P_{p}
\end{equation}
where ${\rho}$ is the surface reflectance of the wall. If the wall had a perfectly smooth surface, we would have:
\begin{equation}
\tag{3}
P_{i}=\rho P_{i}+P_{p}
\end{equation}
Combining (2) and (3), we evince that the following identity should hold:
\begin{equation}
\tag{4}
S^{2}+R^{2}=1
\end{equation}

Equation (4) makes evident that the higher the diffuse power, the lower the specular power. It is worth noting that the power balance in (2) can be considered as global, i.e., it is valid for the whole surface area, if the transmitter is in the Fraunhofer far-field region of the surface and the surface is flat. In this case, in fact, the incident wave can be assumed to be a plane wave, and the angle of incidence can be assumed to be the same over the whole surface. Given the power balance constraint in (2), the contribution of the diffuse scattering field is computed either globally (in the far-field case) or for each tile (in the near-field case), according to a given scattering pattern, as detailed in \cite{ref20}.

\subsection{The Modified Effective Roughness Model for RISs}
\label{subsec2C}
If an ordinary surface is replaced by an RIS, the ER model needs to be generalized. In fact, an RIS is designed to intelligently reradiate the incident wave into the desired direction while minimizing the specular reflection and the diffuse scattering. Therefore, the anomalous reradiation needs to be included into the power balance formulation.

In practice, anomalous reradiation is realized through an appropriate patterning of the surface of the RIS, e.g., by using strip-based, patch-based, or loop-based unit cells. The patterning imposes a spatial modulation on the incident electromagnetic waves, which in turn results in the creation of reradiated modes. The proposed approach for modeling the reradiation from the RIS is, however, macroscopic, i.e., it is not intended for modeling the specific microscopic patterning of the surface. The proposed model characterizes the phase and  amplitude modulation that the RIS applies to the incident electromagnetic waves. Specifically, the generic mode that is reradiated by the RIS is characterized through local surface-averaged (on the scale of the  wavelength) phase and amplitude modulation coefficients, which are denoted by  $\chi \left(x',y'\right)$ and $A\left(x',y'\right)$, respectively, with $\left(x',y'\right)$ being a generic point of the RIS. Further details are given in Section~\ref{sec3}.

Similar to the original ER model, the power balance is imposed to each surface element \textit{dS}. Specifically, the power balance in (1) is generalized to an RIS-coated wall as follows:
\begin{equation}
\tag{5}
P_{i}=P_{r}+P_{s}+P_{m}+P_{d}
\end{equation}
where the power that penetrates into the RIS (\textit{P}$_{p}$ in (1)) is split into two contributions: \textit{P}$_{m}$, which is the total (for all anomalous modes) power that is reradiated by the RIS, and \textit{P}$_{d}$, which is the power that is dissipated into the structure of the RIS. As detailed in Section~\ref{sec3}, each tile of the RIS is viewed as a secondary source of a set of reradiated spherical wavelets, and the superposition of these wavelets results in the reradiated waves based on Huygens' principle.

To characterize the power balance at the RIS surface, we introduce the total reradiation intensity coefficient \textit{m} that  determines the fraction of the  incident power \textit{P}$_{i}$ that is reradiated into the anomalously reradiated modes. In simple terms, \textit{m} plays for anomalous reradiation the same role as the reflectivity ${\rho}$ plays for specular reflection. In RIS-coated walls, in addition, diffuse scattering may be present, which is caused by the presence of possible imperfections.

Under these assumptions and similar to (2), the power balance in (5) can be rewritten as follows:
\begin{equation}
\tag{6}
P_{i}=R^{2}\rho P_{i}+S^{2}P_{i}+R^{2}mP_{i}+\tau P_{i}
\end{equation}
where the same notation as in (2) is used, with the caveat that the Rayleigh factor is applied to both the specular and the reradiated modes and that the coefficient $S^{2}$ is redefined as the ratio between the diffuse power and the incident power. Also, the dissipated power is conveniently expressed as a function of the incident power by using the dissipation parameter ${\tau}$. Equivalently, (6) can be written as follows:
\begin{equation}
\tag{7}
1=R^{2}\rho +S^{2}+R^{2}m+\tau
\end{equation}
If the RIS is assumed to be perfectly smooth and without imperfections (while still taking the power dissipated in the RIS  into account), we obtain:
\begin{equation}
\tag{8}
1=\rho +m+\tau
\end{equation}
Combining (7) and (8), the following identity is established:
\begin{equation}
\tag{9}
S^{2}=\left(1-R^{2}\right)\left(\rho +m\right)
\end{equation}

Even though not explicitly shown in (6), it is worth mentioning that the triplet of parameters $(\rho ,m,\tau )$ depends, in general, on the angle of incidence of the electromagnetic waves, i.e., $(\rho (\theta _{i}),m(\theta _{i}),\tau (\theta _{i}))$. If the RIS is illuminated by several plane waves from different directions, this implies that each signal needs to fulfill (6) based on the corresponding angle of incidence. In wireless communications, this scenario corresponds to a typical multipath propagation channel in which different incident multipath components are scattered by the RIS. In these cases, a complete angle-dependent characterization of the triplet $(\rho (\theta _{i}),m(\theta _{i}),\tau (\theta _{i}))$ is needed, which is usually obtained through full-wave numerical simulations or measurements in an anechoic chamber. If the RIS is a periodic surface, the impact of the angle of incidence can be retrieved by using Floquet’s theory and the mode-matching approach \cite{ref19}.

Under the assumption of far-field illumination, i.e., the incident wave is a plane wave, the aforementioned parameters do not depend on the position of the tile \textit{dS} on the surface. This is similar to the original ER model. In this case, the power balance constraint in (9) can be applied to the whole RIS. If the transmitter is in the near-field region of the RIS, on the other hand, the angle of incidence $\theta _{i}$ depends on the position of the tile on the RIS. Therefore, the power balance in (9) holds only locally, i.e., for the specific tile under consideration, and the macroscopic parameters $(\rho ,m,\tau )$ depend on the location of the tile and the angle of incidence $\theta _{i}$. Some RISs may be realized, at the microscopic scale, by exciting evanescent (i.e., non-propagating) waves in the close proximity of the surface in order to obtain wave transformations at high power efficiencies. These designs result in engineered surfaces with power exchanges between different surface areas of the RIS, where local power losses and local power gains are observed. At the macroscopic level, these implementations of RISs can be modeled through the surface-averaged  amplitude modulation coefficient $A\left(x',y'\right)$ (further detailed are given in Section III).

In (6), the reradiated power can be further expressed as a function of the ensemble of anomalous reradiated modes that are excited by the incident electromagnetic field and that are determined by the physical implementation and specific microstructure of the RIS. As an example, the authors of \cite{ref24} have shown that a phase-gradient RIS that is engineered to operate as an anomalous reflector for a large deflection angle may reradiate power towards three dominant propagating modes: the direction of specular reflection, the desired direction of reradiation, and the direction symmetric to the desired direction of reradiation.

For generality, we assume that the RIS reradiates \textit{N} propagating modes. By denoting with $m_{n}$ the reradiated power coefficient of the \textit{n}$^{th}$ propagating mode, (7) and (9) can be rewritten as follows:
\begin{equation*}
1=R^{2}\rho +S^{2}+R^{2}\sum\nolimits _{n}m_{n}+\tau
\end{equation*}
\begin{equation}
\tag{10}
\implies S^{2}=\left(1-R^{2}\right)\left(\rho +\sum\nolimits _{n}m_{n}\right)
\end{equation}
where $\sum _{n}m_{n}$ can be interpreted as the (macroscopic) power reradiation coefficient of the RIS and the \textit{N}-tuple $\left(m_{1},m_{2},\ldots ,m_{N}\right)$ defines how the reradiated power is distributed among the \textit{N} modes. The \textit{N} modes in (10) do not include the specularly reflected mode, which is accounted for separately by the coefficient ${\rho}$. This is convenient because specular reflection is usually the most significant mode among all parasitic diffracted modes, and it can be efficiently simulated by using conventional ray-based methods.

The focus of this paper is the analysis and modeling of reflective surfaces. The proposed approach can be generalized for application to transmissive surfaces, i.e., RISs that scatter the incident signals towards the forward direction, beyond the wall. In this case, the forward-ER scattering model can be applied \cite{ref26}. The study of this case is left to a future work.

\section{Reradiated Electromagnetic Field}
\label{sec3}

The power balance constraint in (10) ensures that every scattering component generated by an RIS is consistently considered. In this section, we focus our attention on the anomalously reradiated modes and discuss how to account for the spatial modulation along the surface (in phase and amplitude), which results in specified wave transformations.

According to Huygens' principle, as mentioned, each surface element of a finite-size RIS is viewed as a secondary source of a spherical reradiated wavelet, with a given phase and intensity, and the coherent superposition of the wavelets generated by the tiles that comprise the RIS results in the overall reradiated wave. As summarized in Table~I, several methods have been proposed, each one having its own advantages, limitations, and assumptions, to calculate the field reradiated by a finite-size RIS. In the next two sub-sections, we propose two methods: (i)  an integral formulation based on the induction equivalent theorem  \cite{ref22} and on a generalized version of the method of image currents applied to an RIS modeled as an impedance boundary \cite{ref5,ref23}; and (ii) a method based on the antenna theory. The two methods are compared against each other, in order to assess their applicability and performance.

Before introducing the two methods, we summarize in Algorithm~I the proposed approach for computing the complete scattered field from a finite-size RIS, which encompasses the integration of the reradiation model discussed in this section into ray-based models. More specifically: (i) the specular reflected field is obtained by using the geometrical optics methods; (ii) the edge diffracted field related to the specularly reflected field is obtained by applying the uniform theory of diffraction \cite{ref25};  (iii) the diffuse scattering is obtained from the  ER model introduced in Section~\ref{subsec2B} and iv) the anomalously reradiated field is obtained through the methods described in Sections III-A or III-B. The first three contributions are already available in ray-based simulators, and they can then be used in Algorithm~I provided that the power balance constraint in (10) is fulfilled. As mentioned, the focus of the rest of this section is, on the other hand, the computation of the reradiated field, which is not available in current ray-based simulators.

\begin{table}
\begin{tabularx}{0.49\textwidth}{|X|} \hline
\raggedright\arraybackslash{}\textbf{\textit{Algorithm I: Computation of the RIS-scattered total field}}  \\\hline
\raggedright\arraybackslash{}1. Electromagnetic characterization of the RIS (via analysis, full-wave simulations, or measurements). For every reradiated mode, the parameters $\left(\rho ,m_{n},S, \tau \right)$ are defined as a function of the angle of illumination. In this step, the RIS is assumed to be of infinite extent.\\ \hline
\raggedright\arraybackslash{}2. Computation of the field due to specular reflection: $\mathrm{E}_{\text{specular}}$. This is obtained  based on geometrical optics \cite{ref25}. \\\hline
\raggedright\arraybackslash{}3. Computation of the diffracted field from the edges of the RIS, which is related to specular reflection: $\mathrm{E}_{\mathrm{UTD}-\text{diffraction}}$. This is obtained through ray-based models that use, e.g., the uniform theory of diffraction \cite{ref25}. \\\hline
\raggedright\arraybackslash{}4. Computation of the field due to diffuse scattering: $\mathrm{E}_{\text{diffuse}}$. This is obtained through the ER model introduced in Section~\ref{subsec2B}, which is already available in ray-based simulators \cite{ref25}. \\\hline
\raggedright\arraybackslash{}5. Computation of the field reradiated by the RIS: E$_{m}$. This is obtained by using the method introduced in Section~\ref{subsec3A} (see (16)) or in Section~\ref{subsec3B} (see (26) and (31)). \\\hline
\raggedright\arraybackslash{}6. Computation of the total scattered field through coherent summation: $\mathrm{E}_{\mathrm{RIS}}=\mathrm{E}_{\text{specular}}+\mathrm{E}_{\mathrm{UTD}-\text{diffraction}}+\mathrm{E}_{\text{diffuse}}+\mathrm{E}_{m}$ \\\hline
\end{tabularx}
\captionsetup{font=footnotesize,labelfont=footnotesize}
\caption*{Algorithm I. Proposed approach for modeling the total scattered field from a finite-size RIS and its integration into ray-based models and simulators.\vspace{-0.5cm}}
\end{table}

\subsection{Reradiated Field -- Integral Formulation}
\label{subsec3A}

The first proposed method for computing the reradiated field is based on an integral formulation that originates from the induction theorem \cite{ref22,ref23} and a generalization of the method of image currents.

First, we introduce the macroscopic spatial modulation coefficient as:
\begin{equation}
\tag{11}
\Upgamma \left(x',y'\right)=R\cdot \sqrt{m}\cdot A_{m}\left(x',y'\right)\exp \left(j\chi _{m}\left(x',y'\right)\right)
\end{equation}
where $P'=\left(x',y'\right)\in S_{RIS}$ is a generic point of the surface $S_{RIS}$ of the RIS. As introduced in Section~\ref{sec2}, \textit{m} is the reradiation intensity coefficient, and \textit{R} is the Rayleigh factor. Also, $A_{m}\left(x',y'\right)$ and $\chi _{m}\left(x',y'\right)$ are the amplitude and phase of the spatial modulation introduced by the RIS for realizing the desired wave transformation, respectively. To ensure that the power balance constraint in (7) is fulfilled, we assume:
\begin{equation*}
\frac{1}{S_{RIS}}\int _{S_{RIS}}A_{m}^{2}\left(P'\right)dP'=1
\end{equation*}
The surface-averaged coefficient in (11) needs to be interpreted as a macroscopic spatial modulation applied to the incident signal. In other words, $\Upgamma \left(x',y'\right)$ is determined by the actual microscopic structure of the RIS but hides it for analytical tractability. If the RIS reradiates multiple propagating modes, (11) is generalized to:
\begin{equation}
\tag{12}
\Upgamma \left(x',y'\right)=\sum \nolimits _n\Upgamma _n=R \sum \nolimits_n\sqrt{m_{n}} A_{{m_{n}}}\left(x',y'\right) \exp \left(j\chi _{{m_{n}}}\left(x',y'\right)\right)
\end{equation}

If the multiple reradiated modes are plane waves, the amplitude modulation coefficients $A_{{m_{n}}}\left(x',y'\right)$ are constant terms, i.e., $A_{{m_{n}}}\left(x',y'\right)= A_{{m_{n}}}$.

\begin{figure}
\centering
\includegraphics[width=0.4\textwidth]{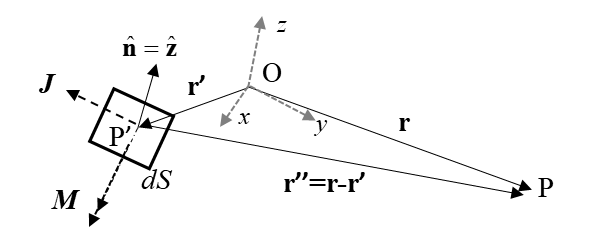}
\caption{Generalized current source that corresponds to a generic surface element \textit{dS}.}
\label{fig:3} \vspace{-0.5cm}
\end{figure}

Let us assume that the macroscopic model of the RIS exemplified by $\Upgamma \left(x',y'\right)$ in (11) and (12) is known, e.g., through some analytical models, full-wave simulations, or measurements. Let us adopt the notation illustrated in Fig.~\ref{fig:3} for a generic surface element (\textit{dS}) of the RIS, with \textit{P} being the observation point where the electromagnetic field is evaluated. More specifically, we consider a finite-size RIS that lies in the $xy(z=0)$ plane of an $Oxyz$ coordinate system. The generic surface element  \textit{dS}  is centered at the point $P'=\left(x',y'\right)\in S_{RIS}$, and $\hat{\mathbf{n}}$ is the normal unit vector that points outwards (i.e., towards the reflection space) the surface element \textit{dS}.

In order to derive the reradiated  field, the induction theorem is applied \cite{ref22,ref23}. The original scattering problem is turned into an equivalent problem (``induction equivalent''), where the sources of the incident field are removed, and a distribution of electric and magnetic current densities, which are the equivalent sources of the reradiated field, is impressed on the RIS surface.

In conventional electromagnetic scattering problems, the induction equivalent problem is usually formulated and solved for surfaces that can be modeled as a perfect electric conductor (PEC) or as a perfect magnetic conductor (PMC). In these cases, the method of image currents is applied, and the total surface current densities are computed. In these cases, either only magnetic or only electric current densities exist. With the current densities at hand, the reradiated field is computed by using the radiation integrals in the absence of the surface. This approach is, however, not directly applicable to engineered surfaces like an RIS. An RIS can be viewed as a layer with a generic surface impedance,  as discussed in \cite{ref18,ref27}. Therefore, both electric and magnetic surface currents need to co-exist simultaneously on its surface. Indeed, a PEC boundary, for example, cannot produce reflections with any phase except $180^\circ$. Therefore, the method of current images needs to be generalized. For brevity, the generalization is detailed in Appendix~A.

Based on Appendix A, the field reradiated by an RIS can be found as generated by equivalent surface electric (\textbf{J}) and  magnetic (\textbf{M}) currents, which depend on the incident fields and on the macroscopic surface modulation coefficient $\Upgamma\left(P'\right)$,  as follows:
\begin{equation}
\tag{13}
\begin{array}{l}
\mathbf{J}\left(P'\right)=\left(1+\Upgamma \left(P'\right)\right)\left[\mathbf{H}_{i}\left(P'\right)\times \hat{\mathbf{n}}\right] \\
\mathbf{M}\left(P'\right)=\left(1-\Upgamma \left(P'\right)\right)\left[\hat{\mathbf{n}}\times \mathbf{E}_{i}\left(P'\right)\right]
\end{array}
\end{equation}
where $\mathbf{E}_{i}$ and $\mathbf{H}_{i}$ denote the incident electric and magnetic fields that are evaluated on the RIS surface. If the RIS operates as a conventional PEC (i.e., $\Upgamma =-1$) we obtain $\mathbf{J}\left(P'\right)=0$ in (13), i.e., the induced electric currents are shorted by the PEC surface. Similar considerations apply if the RIS is configured as a conventional PMC, i.e., $\Upgamma =1$, which implies $\mathbf{M}\left(P'\right)=0$.

Equations (13) are obtained under the assumption that the ratio of the amplitudes of the tangential components of the reflected electric and magnetic fields is the same as the ratio of the amplitudes of the tangential components of the incident electric and magnetic fields. This is the physical optics approximation, i.e., the incident wave is reflected specularly  at every ``point'' $\left(x',y'\right)$ of the RIS. The assumption of locally specular reflection is approximately valid if the surface properties vary slowly at the wavelength scale. This approximated approach is useful in practice, since accurate solutions for the reradiated field exist only for infinite periodical surfaces that are illuminated by plane waves. On the other hand, ray-based algorithms are designed for spherical waves and for finite-size surfaces, which may or may not be periodical. If a periodical RIS is considered, however, the modeling assumption of locally specular reflections implies that the coefficients of the multi-mode expansion in (12) are related but are not exactly the same as the coefficients of the expansion in Floquet's harmonics for periodical metasurfaces \cite{ref19}.
As recently reported in \cite{ref_recent} and \cite{ref_recent2}, it is possible to introduce a ``locally anomalous reflection'' model, which assumes that, at every reflection point, the incident spherical wave is approximated as a plane wave tangential to the wave front, and the reflected field is modeled as an anomalously reflected plane wave. This approach, that describes the RIS  as a locally periodical structure whose period is the size of its constituent super-cell, is approximated as well, since it utilizes the notion of local reflection coefficient for slowly-modulated (at the wavelength scale) metasurfaces.

From (13), the reradiated field can be computed by using the radiation integrals, which account for the finite size and shape of the RIS. This approach, however, ignores the perturbations of the macroscopic modulation coefficient close to the  edges of the RIS. 
This approximation has the same physical ground as the common physical optics approximation: it is acceptable if the size of the RIS is large compared with the wavelength and if the spatial variations of the surface currents are sufficiently slow at the scale of the RIS microstructure.

As a case study, we focus our attention on the field reradiated by an RIS under the assumption that the incident signal is a far-field (with respect to the primary source antenna) spherical wave whose electric and magnetic fields can be formulated as follows:
\begin{align*}
\tag{14}
\mathbf{E}_{i}&=\sqrt{\frac{\eta }{2\pi } P_{t} G_{t}}\cdot e^{j{\chi _{0}}}\cdot \frac{e^{-jk{r_{i}}}}{r_i}\hat{\mathbf{p}}_{i}\\
\mathbf{H}_{i}&=\frac{1}{\eta }\hat{\mathbf{k}}_{i}\times \mathbf{E}_{i}
\end{align*}
where ${\eta}$ is the free-space impedance, \textit{P}$_{\mathrm{t}}$ and \textit{G}$_{\mathrm{t}}$ are the radiated power and the antenna gain of the transmitter, respectively, $\hat{\boldsymbol{p}}_{i}$ is the normal unit vector that embodies the polarization of the incident wave, $\hat{\boldsymbol{k}}_{i}$ is the propagation unit vector, $k=2\pi /\lambda $ is the wave number, and $\chi _{0}$ is a fixed phase shift. Also, $r_{i}$ is the distance from the phase center of the transmitter to the point where the fields are computed. If the fields are observed on the surface of the RIS, then $r_{i}=r_{i}\left(P'\right)$.
It is worth noting that only the tangential components of the fields $\mathbf{E}_{i\tau }$ and  $\mathbf{H}_{i\tau }$, which can be formulated as:
\begin{equation}
\tag{15}
\begin{array}{l}
\mathbf{E}_{i\tau }\left(P'\right)=\hat{\mathbf{n}}\times \left(\mathbf{E}_{i}\left(P'\right)\times \hat{\mathbf{n}}\right)\\
\mathbf{H}_{i\tau }\left(P'\right)=\hat{\mathbf{n}}\times \left(\mathbf{H}_{i}\left(P'\right)\times \hat{\mathbf{n}}\right)
\end{array}
\end{equation}
contribute to the surface currents in (13).

As illustrated in Fig.~\ref{fig:3}, $\boldsymbol{r}'=\left(x',y',z'=0\right)$ denotes the coordinates of the generic point $P'$ of the RIS, $\boldsymbol{r}=\left(x,y,z\right)$ denotes the coordinates of the observation point $P$ whose corresponding normal unit vector is $\hat{\boldsymbol{r}}$, and $\boldsymbol{r}''$ is the difference vector defined as $\boldsymbol{r}''=\boldsymbol{r}-\boldsymbol{r}'$. Assuming that the equivalent surface currents in (13) are the sources of the reradiated fields, the electric field reradiated by the RIS can be formulated as follows (the analytical details are available in Appendix A):
\begin{equation}
\tag{16}
\begin{split}
\mathbf{E}_{\mathrm{m}}\left(P\right)&=\iint _{S_{RIS}}j\frac{e^{-jkr''}}{\lambda r''}\left[\hat{\mathbf{r}}''\times \left(\eta \hat{\mathbf{n}}\times \mathbf{H}_{a}\left(x',y'\right)\right)\times \hat{\mathbf{r}}''\right]dS\\
&+\iint _{S_{RIS}}j\frac{e^{-jkr''}}{\lambda r''}\left[\hat{\mathbf{r}}''\times \left(\mathbf{E}_{a}\left(x',y'\right)\times \hat{\mathbf{n}}\right)\right]dS
\end{split}
\end{equation}
where $r''=\left| \boldsymbol{r}''\right| $, $\hat{\boldsymbol{r}}''=\boldsymbol{r}''/\left| \boldsymbol{r}''\right| $, and the following surface electric and magnetic fields are defined:
\begin{equation}
\tag{17}
\begin{array}{l}
\mathbf{E}_{a}\left(P'\right)=-\dfrac{\left(1-\Upgamma \left(P'\right)\right)}{2}\mathbf{E}_{i\tau }\left(P'\right)\\[8pt]
\mathbf{H}_{a}\left(P'\right)=\dfrac{\left(1+\Upgamma \left(P'\right)\right)}{2}\mathbf{H}_{i\tau }\left(P'\right)
\end{array}
\end{equation}

The fields in (17) can be interpreted as an \textit{RIS-modified Huygens' field source}, and are completely determined by the tangential components of the incident fields and by the macroscopic spatial modulation coefficient in (11) and (12).

It is worth noting that, according to the induction theorem, (16) gives a valid result only if the reradiated field is computed in the space above the surface. Moreover, due to the approximations used in the derivation, [see (A.6) in Appendix~A], (16) is valid if \textit{P} is located in the far-field zone or in the radiative near-field region of the RIS. It is not applicable in the reactive near-field region of the RIS. In practice, it is sufficient that \textit{P} is located at a distance that is a few wavelengths away from the RIS.

The reradiated magnetic field \textbf{H}$_{\mathrm{m}}$ can be obtained, \textit{mutatis mutandis}, by using (13) and then computing the corresponding radiation integral for the magnetic field (see (A.10) in Appendix A).
The formulation in (16) can be generalized to RISs that change the polarization of the incident wave. Due to space limitations, the impact of the polarization is not considered. Simplified closed-form expressions for (16) can be obtained in the Fraunhofer far-field region of the RIS \cite{ref18,ref19} (see Appendix A). Equation (16) is, however, practically relevant since, in some network deployments \cite{Huawei}, the RIS may be large enough that the observation point \textit{P} is located in the radiative near-field region. In general, (16) cannot be expressed in a closed form expression and needs to be computed numerically.

\subsection{Reradiated Field -- Antenna-Array Formulation}
\label{subsec3B}

Based on (16), the field reradiated by a finite-size RIS is formulated in terms of a surface integral. The accuracy of the computation depends on the discretization of the integrand  with the only constraint that the spatial sampling needs to be finer than half of the wavelength $\lambda $. By assuming equal sampling over $x'$ and $y'$, we obtain $dS<\left(\lambda /2\right)\left(\lambda /2\right)=\lambda ^{2}/4$ for ensuring the absence of grating lobes \cite{ref18}. The mentioned discretization is just needed for the numerical calculation of (16) and it is not related to any physical component of the RIS, such as the size of the unit cells. The numerical computation of (16) poses, in general, no problem if a single RIS is considered. \textcolor{black}{If, however, (16) is employed to analyze the system-level performance, e.g., to estimate the coverage maps of a large-scale geographic region in which multiple RISs are deployed, the computational complexity may be considerable.
The issue can be solved, in part, by using parallelized implementations that exploit high-performance graphic processing units (GPUs). In this case, the computation time can be reduced by a factor that is comparable with the number of available GPU cores, i.e., by the order of hundreds in modern GPU architectures \cite{ref21}.
In this sub-section, we discuss another method that can be used in lieu of (16) for reducing the computational complexity and to simplify the analytical formulation}.

In the literature, another approach that has been used for estimating the field reradiated by a finite-size RIS is based on antenna theory. In this method, an RIS is viewed as an array of scattering antenna elements \cite{ref9,ref10,ref11}. Specifically, the RIS is subdivided into surface elements of area $\Delta S$, and each surface element can be thought of as an aperture antenna that receives the incident power \textit{P}$_{i}$ and reradiates a spherical wavelet whose power is \textit{P}$_{m}$, which is defined in (5), according to a given power radiation pattern $f_{\mathrm{m}}\left(\theta _{m}\right)$, where $\theta _{m}$ is the angle of observation of the reradiated wave. In general, $f_{\mathrm{m}}\left(\theta _{m}\right)$ needs to be appropriately chosen. In the following, we first elaborate on the constraints that $\Delta S$ needs to fulfill for ensuring that the antenna-array model for the RIS is physically consistent.

The representation of a finite-size RIS as an antenna-array is based on the known relation between the effective aperture $A_{m}$ and the directivity $D_{m}$ of an aperture antenna, i.e., $D_{m}=\left(4\pi /\lambda ^{2}\right)A_{m}$ \cite{ref23}. If $\Delta S$ is viewed as the physical size of the aperture antenna, the antenna effective aperture needs to be smaller than the size of the physical aperture, i.e., $A_{m}\leq \Delta S$. Otherwise, the antenna could receive and could reradiate more power than the power that enters into the aperture (i.e., $P_{m}>P_{i}$), which is possible only for resonant antenna elements.
In this paper, we assume that the antenna elements are not resonant and behave as electrically large aperture antennas, for which $A_{m}\leq \Delta S$ \cite{ref28}.

To formulate an antenna-array model that is electromagnetically consistent, therefore, we consider antenna elements for which the following physical relations need to hold:
\begin{equation}
\tag{18}
\begin{array}{lll}
\left\{\begin{array}{l}
D_{m}=\left(4\pi /\lambda ^{2}\right)A_{m}\\
A_{m}\leq \Updelta S\\
\Updelta S\leq \lambda ^{2}/4
\end{array}\right. & \Rightarrow & D_{m}\leq \pi \end{array}
\end{equation}
where the third inequality ensures the absence of grating lobes.
For simplicity of notation, we assume $\Delta S=\left(\Delta l\right)^{2}$. From the third equation in (18), we obtain $\Delta l\leq \lambda /2$ and $\Delta l=\delta \lambda $ with $\delta \leq \delta _{\max}=1/2$. Thus, from (18), we have:
\begin{equation}
\tag{19}
\begin{array}{ll}
D_{m} & =\left(4\pi /\lambda ^{2}\right)A_{m}\leq \left(4\pi /\lambda ^{2}\right)\Updelta S\\ & =4\pi \left(\Updelta l/\lambda \right)^{2}=4\pi \delta ^{2}
\end{array}
\end{equation}

By definition, the directivity \textit{D}$_{m}$ is always greater than one. Thus, from (19), we obtain:
\begin{equation*}
\begin{array}{l}
4\pi \delta ^{2}\geq D_{m}\geq 1
\Rightarrow \delta \geq \delta _{\min }=1/\left(2\sqrt{\pi }\right)\approx 0.28
\end{array}
\end{equation*}
Therefore, ${\delta}$ must satisfy the following constraints:
\begin{equation}
\tag{20}
0.28=\delta _{\min}\leq \delta \leq \delta _{\max}=0.5
\end{equation}

From the lower bound in (20), we evince that modeling a finite-size RIS as an antenna-array can be considered correct if the size of each antenna element is greater than $\Delta l_{\min}=0.28\lambda $. A similar finding was recently identified through experimental measurements in \cite{ref9} and \cite{ref10}. By assuming that the surface element $\Delta S$ is chosen equal to the size of the unit cells, the authors of \cite{ref9} and \cite{ref10} remark  that the antenna radiation pattern needs to be chosen as a function of the size of the unit cell as well, in order to obtain physically consistent and accurate results. It is worth mentioning, in addition, that the mutual coupling among the antenna elements cannot be ignored if $\Delta l<\lambda /2$, but it can be approximately taken into account by applying locally periodic boundary conditions when estimating the reflection coefficient of each unit cell \cite{ref9,ref10,ref19}.

Motivated by \cite{ref9} and \cite{ref10}, let us analyze the interplay between the antenna pattern $f_{m}\left(\theta _{m}\right)$ and the size of the surface element $\Delta S$. An example of antenna pattern that is often utilized in the literature is the exponential-Lambertian function (see, e.g., \cite{ref9,ref10}). The corresponding power radiation pattern is:
\begin{equation}
\tag{21}
f_{m}\left(\theta _{m}\right)=\left(\cos \theta _{m}\right)^{\alpha },\quad \theta _{m}\in \left[0,\pi /2\right]
\end{equation}
where $\alpha \geq 0$ is a tuning parameter.
From (21), by definition of directivity, we obtain $D_{m}=2\left(\alpha +1\right)$ \cite[Eq. (16)]{ref10}. From (18), we have (for every $\delta $):
\begin{equation}
\tag{22}
D_{m}=2\left(\alpha +1\right)\leq \pi \Rightarrow \alpha \leq \frac{\pi }{2}-1\approx 0.57
\end{equation}

On the other hand, larger values of $\alpha $ would result in grating lobes. Since, in addition, $\alpha \geq 0$, from (19) we obtain:
\begin{equation}
\tag{23}
4\pi \delta ^{2}\geq \left.2\left(\alpha +1\right)\right| _{\alpha =0}\Rightarrow \delta \geq 1/\sqrt{2\pi }\approx 0.4
\end{equation}

From (22), (23), and (20), we see that the radiation pattern in (21) can be employed for modeling an RIS only if $0.4\leq \delta \leq 0.5$ and $\alpha \leq 0.57$. If the surface element $\Delta S$ is assumed equal to a unit cell of the RIS, as in \cite{ref9} and \cite{ref10}, the exponential-Lambertian radiation pattern cannot be used for RISs whose unit cells are electrically small, if i.e., $\delta < 0.4$.

Another often utilized antenna pattern is the Huygens radiation pattern, which is defined as:
\begin{equation}
\tag{24}
f_{m}\left(\theta _{m}\right)=\left(\left(1+\cos \theta _{m}\right)/2\right)^{2},\quad \theta _{m}\in \left[0,\pi \right]
\end{equation}

The power antenna pattern in (24) is motivated by the fact that the Huygens source constitutes a reference model for small aperture antennas. Thus, it can be considered as a suitable choice for representing a discretized implementation of Huygens' principle \cite{ref28} and for modeling an RIS as an antenna-array.

By definition of directivity, we obtain \textit{D}$_{m}$=3 from (24). From (19), therefore, we have:
\begin{equation}
\tag{25}
4\pi \delta ^{2}\geq D_{m}=3
\Rightarrow \delta \geq \frac{1}{2}\sqrt{\frac{3}{\pi }}\simeq 0.49
\end{equation}

Equation (25) unveils that the Huygens power radiation pattern can be applied only if the size $\Delta S$ of the surface elements of the RIS is greater than or equal to 0.49${\lambda}$.

The analysis of the exponential-Lambertian and Huygens' radiation patterns bring to our attention that modeling an RIS as an antenna-array is possible, but some constraints on the modeling parameters need to be ensured. Specifically, an RIS can be modeled as an antenna-array provided that the feasibility conditions in (18)--(20) are fulfilled. For example, the size $\Delta S$ of the surface elements  and the power radiation pattern are interrelated. As a result, the analysis shows that $\Delta S$ may not be chosen equal to the size of the unit cells of the RIS, if the RIS is made of unit cells whose size is much smaller than $\lambda/2$ and if the exponential-Lambertian or the Huygens radiation patterns for the unit cells are utilized.

Let us assume that the feasibility conditions in (18)--(20) are fulfilled. Then, we are in a position to formulate the field reradiated by an RIS as the sum of the far-field spherical wavelets that are reradiated by each surface element $\Delta S$. This is similar to the approach utilized in Section III-A, with the difference that we consider a discretized version of the RIS and that each surface element $\Delta S$ is associated with a given power radiation pattern $f_{m}\left(\theta _{m}\right)$. Similar to Section III-A, the approach introduced in this section can be applied in the far-field and in the radiative near-field regions of the RIS.

Specifically, the RIS is partitioned into $N_{X}\times N_{Y}$ elementary surface elements $\Delta S$. The generic surface element is identified by the indices $\left(u,v\right)$, where $u=1,2,\ldots ,N_{X}$ and $v=1,2,\ldots ,N_{Y}$. The corresponding reradiated electric field is denoted by $\Updelta \mathbf{E}_{\mathrm{m}}\left({u},{v}\right)$. Then, the total reradiated field evaluated at the observation point $P$ can be formulated as:
\begin{equation}
\tag{26}
\mathbf{E}_{m}\left(P\right)=\sum _{u=1}^{N_{X}}\sum _{v=1}^{N_{Y}}\Updelta \mathbf{E}_{m}\left(P\left| u,v\right.\right)
\end{equation}

To compute $\Updelta \mathbf{E}_{\mathrm{m}}\left({u},{v}\right)$, we use a two-step approach: (i) first, we ensure that the reradiated power fulfills the power balance constraint in (10) and then (ii) we account for the phase/amplitude modulation that each surface element $\Updelta S$ needs to apply in order to realize the desired wave transformation.

As far as the power balance principle is concerned, the quadruplet of parameters (\textit{m,} ${\tau}$\textit{,} ${\rho}$\textit{, S}) needs to fulfill (10). In the proposed approach, the presence of possible non-ideal reradiation effects and losses is accounted for by the parameter $m<1$. This implies that the surface elements can be modeled as ideal scattering aperture antennas and that the effective area \textit{A}$_{m}$ of the antenna can be assumed to be equal to the geometrical  size of the surface element ${\Updelta}$\textit{S}, i.e., \textit{A}$_{m}$=${\Updelta}$\textit{S}. Assuming, e.g., that Huygens' power radiation pattern is utilized for each surface element, this implies ${\Updelta}$\textit{l}=0.49\textit{${\lambda}$}.
As a byproduct, this choice ensures that the mutual coupling among the surface elements may be assumed negligible for the first-order analysis.

Under these assumptions, let $\theta _{i}\left(u,v\right)$ and $\theta _{m}\left(u,v\right)$ denote the direction of the incident wave and the direction of propagation towards the observation point $P$, respectively, that correspond to the surface element $\left(u,v\right)$. Then, the reradiated electric field can be formulated as follows:
\begin{multline}
\tag{27}
\Updelta \mathbf{E}_{m}\left(P\left| u,v\right.\right)=\Updelta E_{m0}\left(P\left| u,v\right.\right)\sqrt{f_{m}\left(\theta _{m}\left(P\left| u,v\right.\right)\right)}\\
\cdot \Upgamma \left(u,v\right)\exp \left(-jk\left(r_{i}\left(u,v\right)+r_{m}\left(P\left| u,v\right.\right)\right)\right)\hat{\mathbf{p}}_{m}
\end{multline}
where $\hat{\boldsymbol{p}}_{m}$ is the unit normal vector that embodies the polarization of the reradiated wave, $\Updelta E_{m0}\left(P|u,v\right)$ is the complex amplitude of the reradiated wave, $\Upgamma \left(u,v\right)$ is the macroscopic spatial modulation coefficient in (11) and (12) that is evaluated at $\left(x',y'\right)=\left(\Updelta l u,\Updelta l v\right)$, and the exponential term accounts for the accumulated phase shift along  the path from the transmitter to the $\left(u,v\right)$th surface element of the RIS and from the latter surface element to the receiver, which depends on the distances $r_{i}\left(u,v\right)$ and $r_{m}\left(P|u,v\right)$.

To fulfill the power conservation principle, it needs to be ensured that a fraction equal to \textit{mR}$^{2}$ of the power received by the antenna element is reradiated into the upper half-space, which is a solid angle of 2${\uppi}$ steradians. Therefore, the following power balance equation needs to hold:
\begin{align*}
mR^{2}\frac{\left| \mathbf{E}_{i}\right| ^{2}}{2\eta }A_{m}\left(\theta _{i}\right)&=mR^{2}\frac{\left| \mathbf{E}_{i}\right| ^{2}}{2\eta }\frac{\lambda ^{2}}{4\pi }f_{m}\left(\theta _{i}\right)\\
&=\int _{2\pi }\frac{\left| \Updelta \mathbf{E}_{m}\left(P\right)\right| ^{2}}{2\eta }r_{m}^{2}d\Upomega =2\pi \frac{\left| \Updelta E_{m0}\left(P\right)\right| ^{2}}{2\eta }\\
&\cdot \int _{0}^{\pi /2}f_{m}\left(\theta _{m}\right)r_{m}^{2}\left(P\right)\sin \left(\theta _{m}\right)d\theta _{m}
\end{align*}
where the dependence on (\textit{u,v}) is omitted for ease of reading.

Therefore, we obtain:
\begin{equation}
\tag{28}
\begin{split}
mR^{2}\left| \mathbf{E}_{i}\right| ^{2}\frac{\lambda ^{2}}{4\pi }f_{m}\left(\theta _{i}\right)&=2\pi \left| \Updelta E_{m0}\left(P\right)\right| ^{2}\\
&\cdot \int _{0}^{\pi /2}f_{m}\left(\theta _{m}\right)r_{m}^{2}\left(P\right)\sin \left(\theta _{m}\right)d\theta _{m}
\end{split}
\end{equation}
where $\left| \boldsymbol{E}_{i}\right| ^{2}=(\eta /2\pi) P_{t}G_{t}/r_{i}^{2}\left(u,v\right)=E_{i1}/r_{i}^{2}\left(u,v\right)$ is the power intensity of the incident electric field according to (14).

As a result, the power intensity of the total reradiated field can be formulated as follows:
\begin{equation}
\tag{29}
\left| \Updelta E_{m0}\left(P\right)\right| ^{2}=\frac{\lambda ^{2}}{8\pi ^{2}}\frac{mR^{2}E_{i1}}{\int _{0}^{\pi /2}f_{m}\left(\theta _{m}\right)\sin \left(\theta _{m}\right)d\theta _{m}}\frac{f_{m}\left(\theta _{i}\right)}{r_{i}^{2}r_{m}^{2}}
\end{equation}

By combining (27) and (29), the reradiated electric field can be formulated as follows:
\begin{equation}
\tag{30}
\begin{split}
\Updelta \mathbf{E}_{m}\left(P\left| u,v\right.\right)&=\frac{\lambda }{2\pi }\frac{\sqrt{m}RE_{i1}}{\sqrt{2\int _{0}^{\pi /2}f_{m}\left(\theta _{m}\right)\sin \left(\theta _{m}\right)d\theta _{m}}}\Upgamma \left(u,v\right)\\
&\cdot \sqrt{f_{m}\left(\theta _{i}\left(u,v\right)\right)}\sqrt{f_{m}\left(\theta _{m}\left(P\left| u,v\right.\right)\right)}\\
&\cdot \frac{\exp \left(-jk\left(r_{i}\left(u,v\right)+r_{m}\left(P\left| u,v\right.\right)\right)\right)}{r_{i}\left(u,v\right)r_{m}\left(P\left| u,v\right.\right)} \hat{\mathbf{p}}_{m}
\end{split}
\end{equation}

Notably, $\theta _{m}\left(P\left| u,v\right.\right)$ in (30) denotes the angle between the $(u,v)$th tile and the observation point $P$. This latter angle is, in general, different for each tile and is different from the desired angle of reflection that is determined by the macroscopic modulation coefficient $\Upgamma$.

Based on (30), the corresponding magnetic field can be obtained by using the local plane wave approximation for the generic surface element of the RIS:
\begin{equation*}
\Updelta \mathbf{H}_{m}=\left(1/\eta \right)\,\,\hat{\mathbf{r}}_{m}\times \mathbf{E}_{i}
\end{equation*}

If the Huygens power radiation pattern is assumed, (30) can be formulated as a closed-form expression, as follows:
\begin{equation}
\tag{31}
\begin{split}
\Updelta \mathbf{E}_{m}\left(P\left| u,v\right.\right)&= \sqrt{m}R\,\,E_{i1}\Upgamma \left(u,v\right)\\
&\cdot \frac{3\lambda }{16\pi }\left(1+\cos \theta _{i}\left(u,v\right)\right)\left(1+\cos \theta _{m}\left(P\left| u,v\right.\right)\right)\\
&\cdot \frac{\exp \left(-jk\left(r_{i}\left(u,v\right)+r_{m}\left(P\left| u,v\right.\right)\right)\right)}{r_{i}\left(u,v\right)r_{m}\left(P\left| u,v\right.\right)} \hat{\mathbf{p}}_{m}
\end{split}
\end{equation}

Finally, the complete reradiated electric field is obtained by inserting (30) in (26). It is worth noting that (30) satisfies the reciprocity condition if the surface-averaged macroscopic coefficient $\Upgamma$ is a reciprocal function of the angle of incidence and the desired angle of reflection. In the next section, the desired angle of reflection is denoted by $\theta_r$.

\section{Model Validation}
\label{sec4}
In this section, we validate the proposed reradiation models for RISs. Specifically, we consider examples of RISs for which the macroscopic spatial modulation coefficient $\Upgamma$ is obtained from analytical models, full-wave simulations, and experimental measurements. Six case studies are analyzed.
\begin{figure}
\centering
\includegraphics[width=0.4\textwidth]{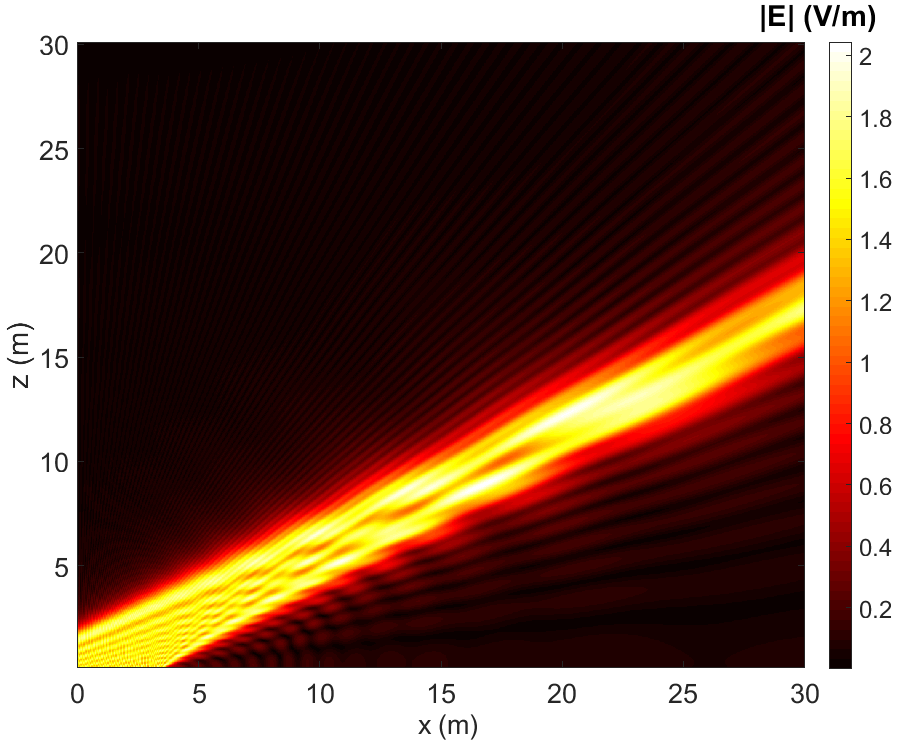}
\caption{Reradiated field [V/m] in the \textit{xz} plane from a $7 \times 7$ m$^2$ RIS that is located in the $xy$ plane and is centered at the origin. Setup: 3 GHz operating frequency, normal incidence, reflection towards the angle of 60 degrees.}
\label{fig:4} \vspace{-0.5cm}
\end{figure}

\subsubsection{Ideal Phase-Gradient Reflector} The first case study corresponds to an ideal metasurface that introduces an ideal phase modulation to the reflected fields, with the goal of reradiating a single incident plane wave towards a target direction in the absence of dissipation and undesired  reradiated modes. We consider a $7 \times 7\:m^2$ large RIS that lies in the \textit{xy} plane and that introduces a linear phase modulation such that $d\chi _{m}\left(x'\right)/dx'=k\left(\mathit{\sin } \theta _{i}-\mathit{\sin } \theta _{r}\right)$, where $\theta _{i}$ is the angle of incidence and $\theta _{r}$ is the desired angle of reflection \cite{ref24}. Being an ideal case, the rest of the parameters in (11) are set equal to $m=1$, $R=1$, and $A_m(x',y')=1$.

The RIS operates at 3 GHz and is illuminated normally ($\theta _{i}=0$) by a plane wave that is linearly polarized in the $y$-direction and whose intensity is 1 V/m. The desired angle of reradiation is $\theta _{r}=60$ degrees. The reradiated field is computed with the integral model in Section \ref{subsec3A} and is reported in Fig.~\ref{fig:4}. The locations illustrated in Fig.~\ref{fig:4} lie in the radiative near-field region of the RIS, since the Fraunhofer far-field distance is approximately equal to 1000 m for the considered setup. We observe that the electric field is steered towards the desired angle of reflection. As expected, in addition, we observe edge-diffraction fringes that are due to the finite size of the RIS.

For comparison, we compute the reradiated electric field by using the antenna-array model introduced in Section \ref{subsec3B}, i.e., by using (26) and (31). For consistency, (16) is computed by using the same discretization as for (26). The comparison is illustrated in Fig.~\ref{fig:5}. The figure shows that the relative error is less than 1-2\% for most of the observation points. Thus, both models provide consistent results. The antenna-array method is, however, simpler and faster to compute. \textcolor{black}{In particular, the computation of the methods described in Section \ref{subsec3A} and in Section \ref{subsec3B} requires 0.13 s and $1.8\cdot10^{-2}$ s, respectively, for each observation point. Therefore, the method in Section \ref{subsec3B} is six times faster than the method in Section \ref{subsec3A}.}
\begin{figure}
\centering
\includegraphics[width=0.4\textwidth]{images/image5.png}
\caption{Relative error (in percentage) of the antenna-array model with respect to the integral formulation in (16).}
\label{fig:5} \vspace{-0.5cm}
\end{figure}

\begin{figure}
\centering
\includegraphics[width=0.4\textwidth]{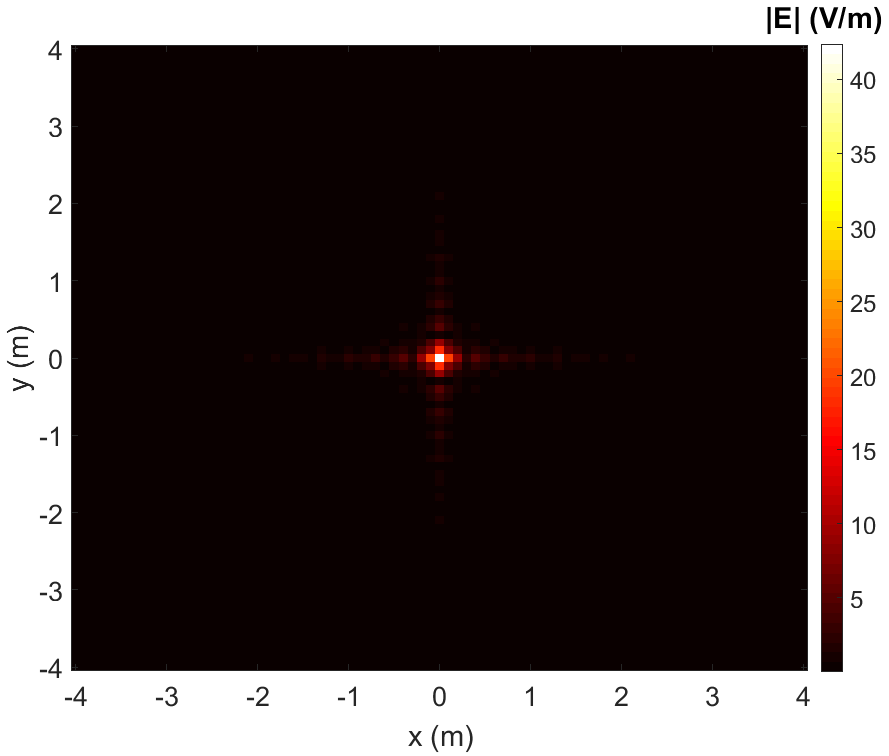}
\caption{Reradiated field [V/m] from a $7 \times 7$ m$^2$ focusing lens. Except for the phase profile of the metasurface, the setup is the same as for Fig.~\ref{fig:4}.}
\label{fig:6} \vspace{-0.5cm}
\end{figure}
\subsubsection{Ideal Focusing Lens} The second canonical case study corresponds to an ideal and lossless metasurface that focuses a single spherical wave towards an intended location, i.e., the metasurface operates as a reflecting focusing lens. In this example,
the angle of incidence is $\theta _{i}=60$ degrees, the RIS lies in the \textit{xy} plane at $z_{0}=-10$~m, and the intended  focus is at the origin. The phase modulation profile to obtain the desired reradiated wave is $\chi _{{m_{1}}}\left(x',y'\right)=k\sqrt{\left(x'\right)^{2}+\left(y'\right)^{2}+\left(z_{0}\right)^{2}}-k\mathit{\sin } \theta _{i}x'$, while we assume unitary amplitude parameters as in the previous case study. The results are illustrated in Fig.~\ref{fig:6}. We see that the electric field is focused at the desired location and that the intensity (ignoring the impact of the transmission distances) is approximately 40 times stronger than the incident field, thanks to the focusing capabilities of the RIS for this considered application case. It is worth noting that the location of the focusing point lies in the radiative near-field of the RIS.

\subsubsection{Lossless Anomalous Reflector (with parasitic modes)} To evaluate the capabilities of the proposed model to account for the presence of parasitic reradiated modes, we consider the metasurface analyzed in \cite[Fig. 9]{ref19}. This case study corresponds to a phase-gradient RIS that is lossless, periodic, and is optimized based on the locally periodic approximation. In \cite{ref19}, the metasurface is characterized with the aid of electromagnetic simulations and on an approximate analytical framework based on Floquet's theory. In this example, the RIS is illuminated by a normally incident plane wave at 3~GHz. The modulation period of the metasurface is $D = 0.1064$~m and the size of the RIS is $10D \times 10D$. The reradiated field is evaluated at a distance equal to 22.64~m, which is close to the Fraunhofer far-field boundary.
\begin{figure}
\centering
\includegraphics[width=0.4\textwidth]{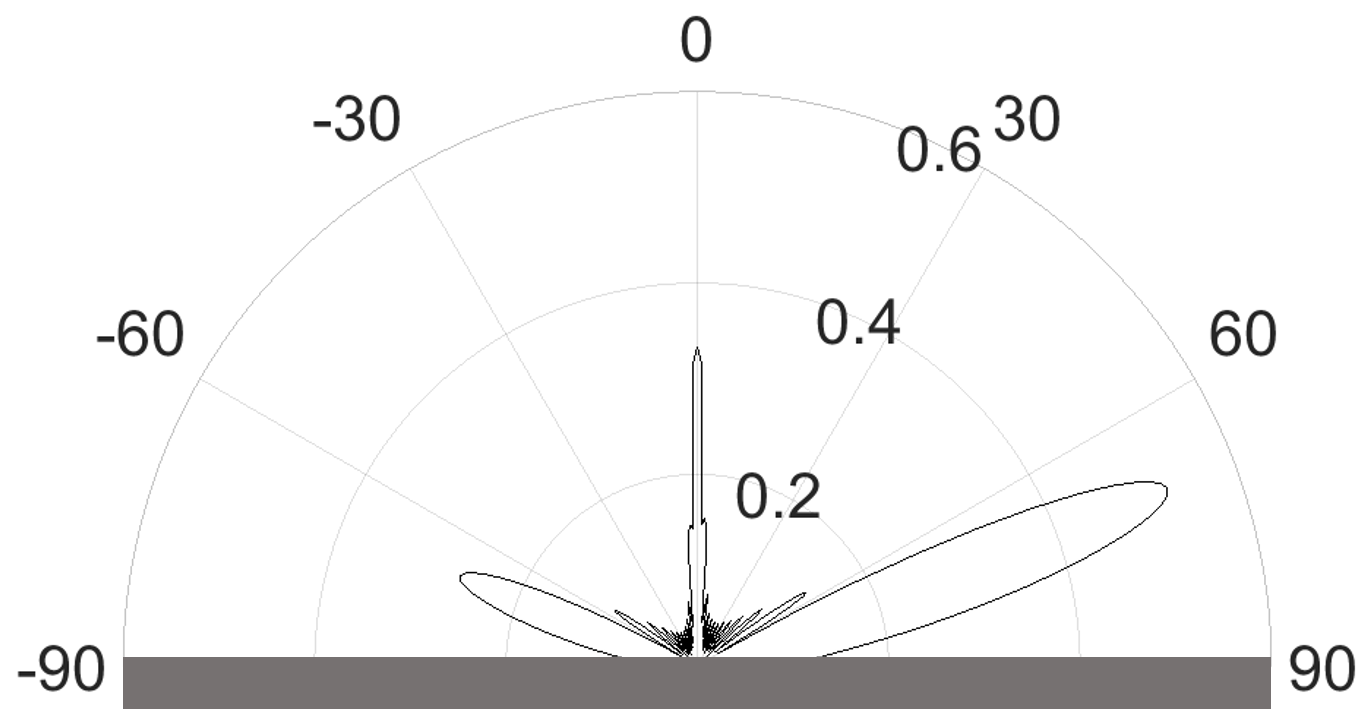}
\caption{Radiation pattern of the lossless anomalous reflector in \cite{ref19}. Setup: normal incidence and the desired angle of reflection is 70 degrees.}
\label{fig:7} \vspace{-0.5cm}
\end{figure}
The reradiation pattern is illustrated in Fig.~\ref{fig:7} and is obtained by using the antenna-array formulation in Section~\ref{subsec3B} in the absence of losses and diffuse scattering, as in \cite{ref19}. The reradiated field in (26) is computed by taking into account two anomalous reradiated modes that are combined with specular reflection and diffraction, according to Algorithm~I. Specifically, specular reflection and diffraction are obtained by using state-of-the-art ray-based methods. Specular reflection, anomalous reradiation, and diffraction are appropriately weighted according to the power balance principle in (10) and are then coherently summed together to obtain the total scattered field. More precisely, Fig.~\ref{fig:7} is obtained by setting $\rho$=0.17 (undesired specular reflection), $m_{1}$=0.76 (desired anomalous reflection), and $m_{2}$=0.17 (undesired symmetric reflection). By comparing Fig.~\ref{fig:7} with \cite[Fig. 9]{ref19}, we observe a good agreement between the two reradiation patterns.

\subsubsection{Diffuse Scattering} 
\begin{figure}
\centering
\includegraphics[width=0.4\textwidth]{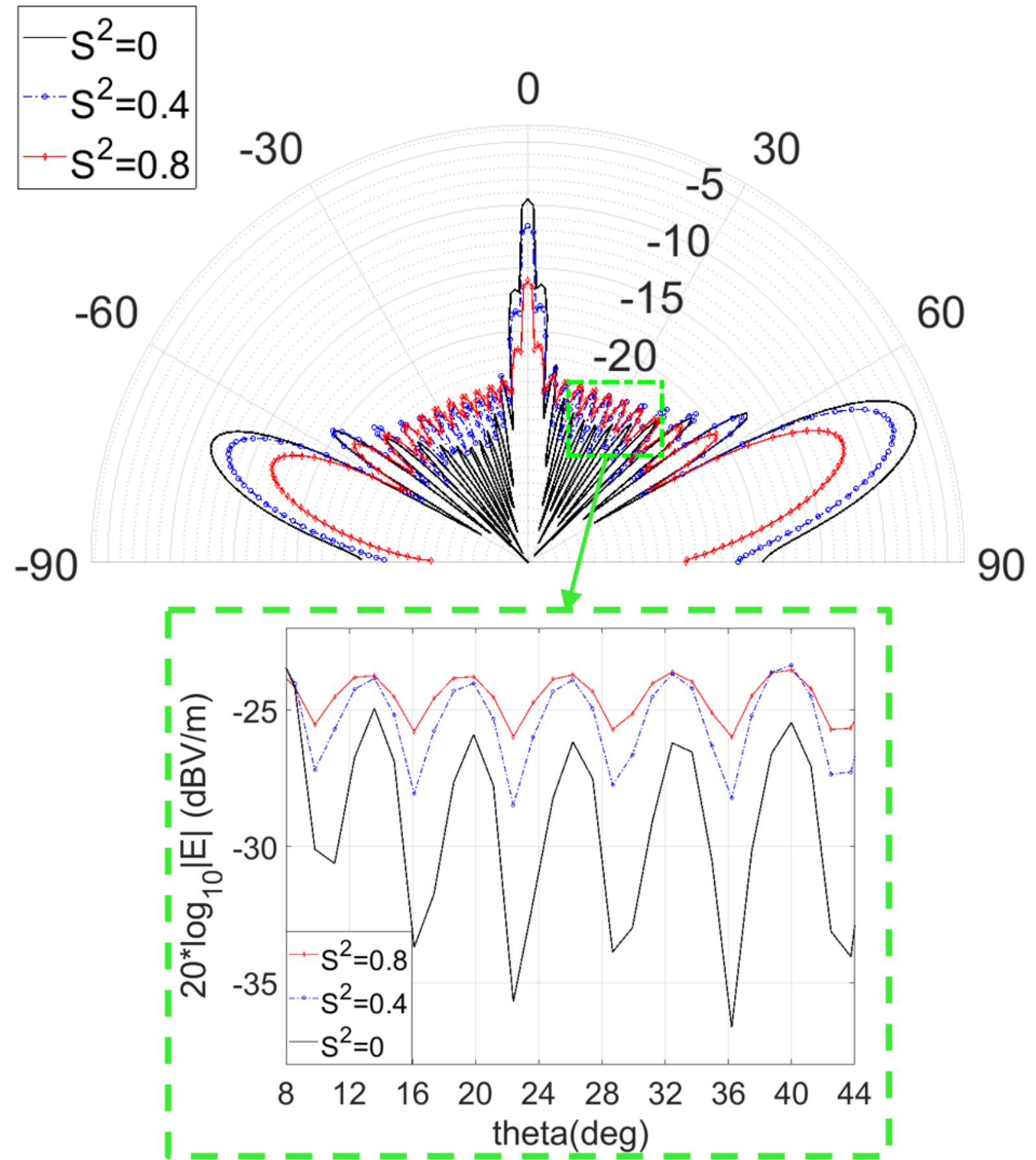}
\caption{Far-field (dBV/m) scattering pattern of the lossless anomalous reflector considered in \cite{ref19} in the absence of diffuse scattering (reference black curve), and in the presence of 40\% (blue curve) and 80\% (red curve) of the incident power diverted into diffuse scattering.}
\label{fig:8} \vspace{-0.5cm}
\end{figure}

The fourth case study is centered on analyzing the impact of diffuse scattering that originates from design trade-offs, construction nonidealities and/or the deposit of dust on the surface of the RIS. To the best of our knowledge, no specific experimental results on modeling diffuse scattering from engineered surfaces exist in the literature. Thus, we conduct a parametric study, in order to assess the potential impact of diffuse scattering on the radiation pattern of an RIS. The numerical results are shown in Fig.~\ref{fig:8} for different values of the scattering parameter \textit{S} in (10). The metasurface considered in Fig.~\ref{fig:8} is the lossless anomalous reflector analyzed in Fig.~\ref{fig:7}. In Fig.~\ref{fig:8}, we study the impact of diffuse scattering under the assumption that 40\% (\textit{S}$^{2}$=0.4) or 80\% (\textit{S}$^{2}$=0.8) of the incident power is diverted into Lambertian diffuse scattering according to the power balance constraint in (10). Although the considered values for $S$ may be overestimated, we choose them to make the curves more readable. Figure~\ref{fig:8} shows the important role that diffuse scattering can play in RIS-aided communications. In particular, we see that the intensity of the electric field towards the desired direction of reflection is reduced, and that the intensity of the sidelobes increases as \textit{S} increases. Differently from Fig.~\ref{fig:7}, a logarithmic scale is used to highlight the sidelobes.
\subsubsection{Lossy Anomalous Reflector} 
\begin{figure}
\centering
\includegraphics[width=0.4\textwidth]{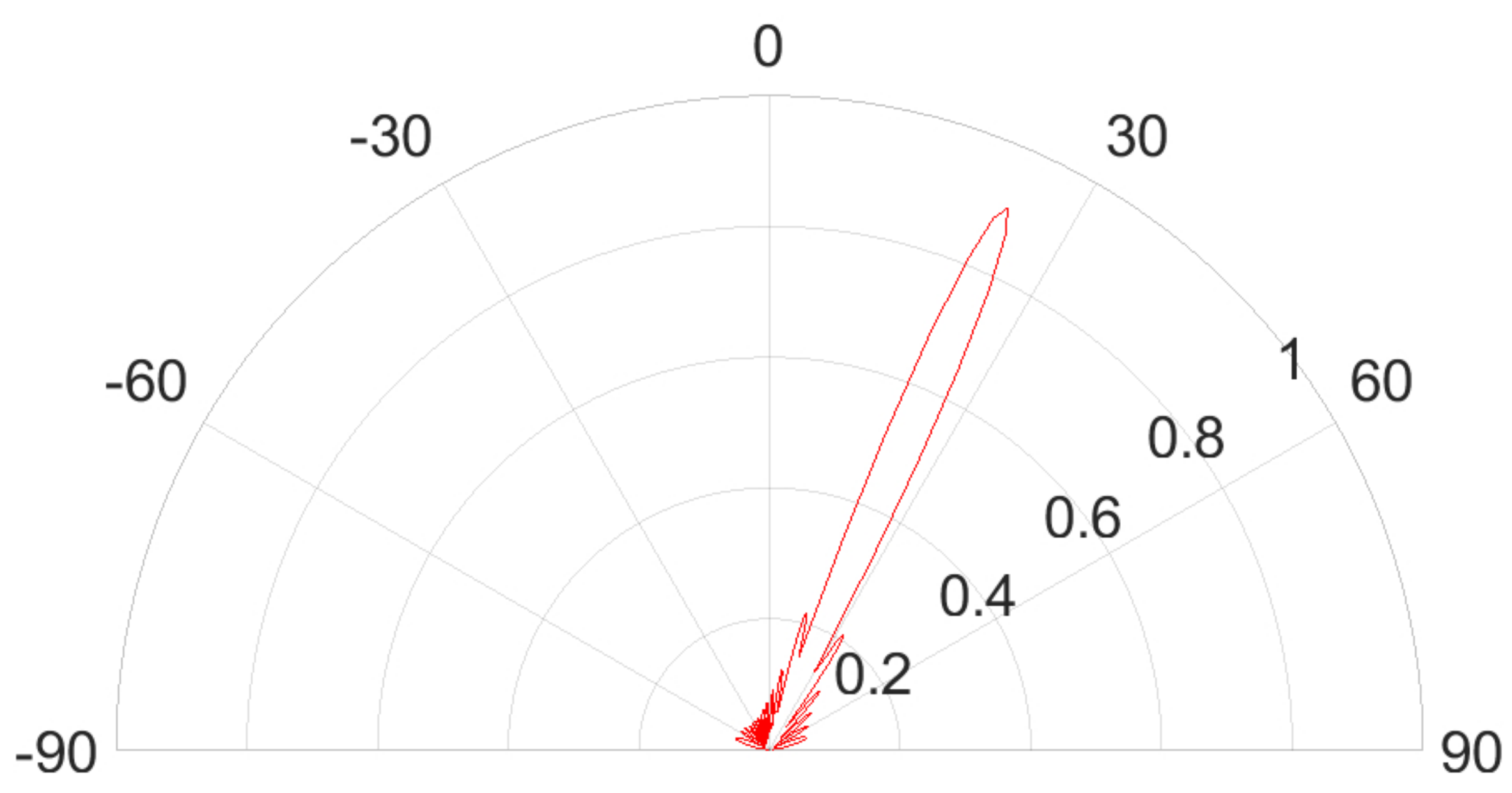}
\caption{Far-field radiation pattern ($\rho =0$, $S=0, m=0.97$) corresponding to the lossy phase-gradient metasurface reported in \cite[Fig~2.c]{ref29}. The corresponding benchmark radiation pattern is available in \cite[Fig~2.d]{ref29}.}
\label{fig:9} \vspace{-0.5cm}
\end{figure}

\begin{figure}
\centering
\includegraphics[width=0.4\textwidth]{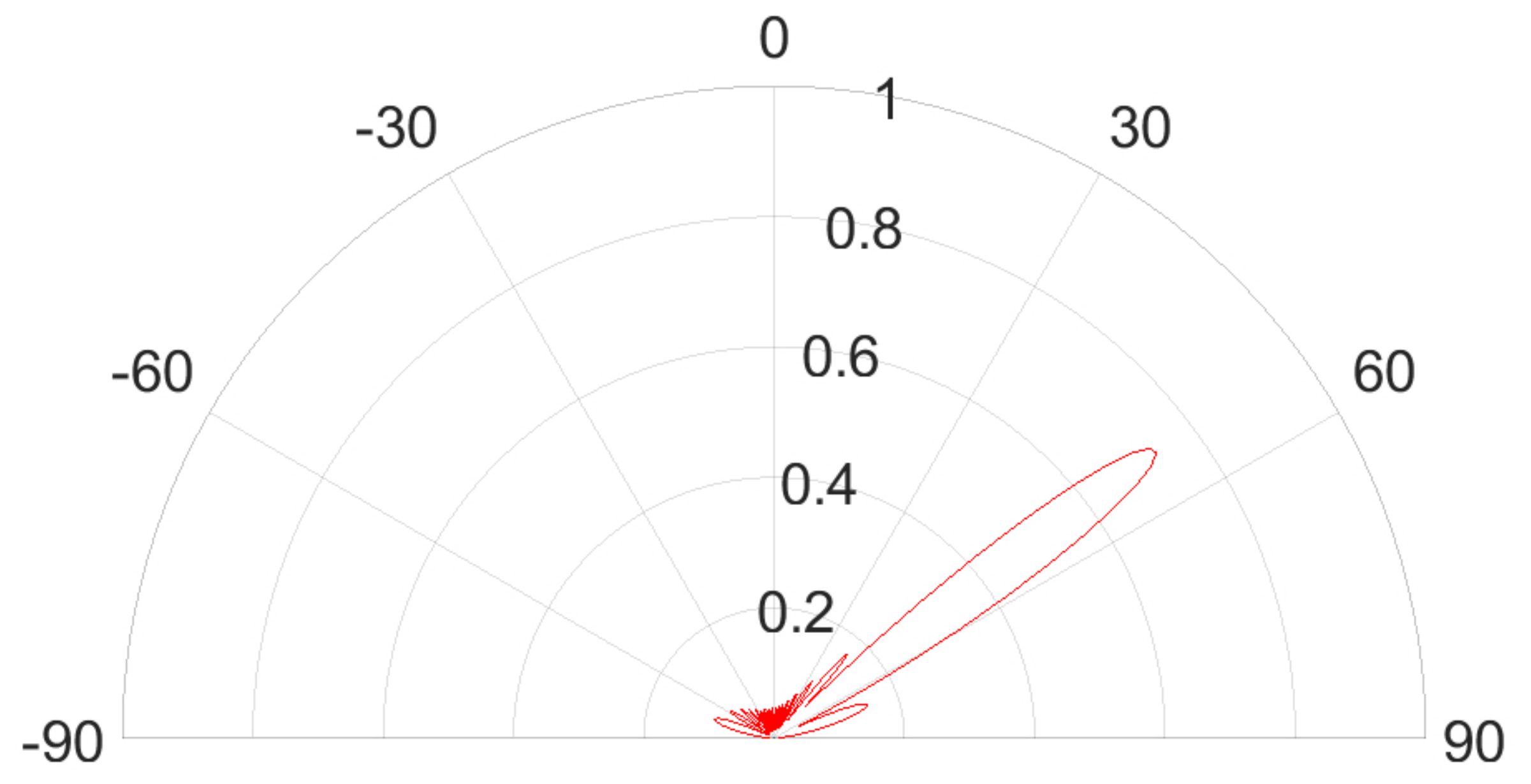}
\caption{Far-field radiation pattern ($\rho =0$, $S=0, m=0.9$) corresponding to the lossy phase-gradient metasurface reported in \cite[Fig~2.e]{ref29}. The corresponding benchmark radiation pattern is available in \cite[Fig~2.f]{ref29}.}
\label{fig:10} \vspace{-0.5cm}
\end{figure}

The fifth case study is centered on validating the proposed macroscopic model for characterizing a  phase-gradient lossy metasurface that has been manufactured and experimentally  characterized  \cite{ref29}. To match the parameters of the proposed model with the metasurface designed in \cite{ref29}, we set $\rho =0$ and $S=0$. In addition, the metasurface in \cite{ref29} is designed to realize anomalous reflection and to suppress the parasitic reradiation modes.
Although the metasurface is engineered based on the local design, for moderate angles of reradiation, the scattering into parasitic modes can be ignored. Part of the incident power is dissipated, but a single reradiated mode exists, and (10) reduces to $m+\tau =1$. The rest of the simulation parameters are the same as those in \cite{ref29}. The reradiation pattern is obtained by using the antenna-array model and is illustrated in Fig.~\ref{fig:9} and Fig.~\ref{fig:10} for the desired angles of reflection, and by setting $m=0.97$ and $m=0.9$, respectively.
From Fig.~\ref{fig:9} and Fig.~\ref{fig:10}, we see that the reradiation patterns are in good agreement with those reported in \cite{ref29}. This confirms the suitability and accuracy of the proposed macroscopic reradiation and power conservation models.

\subsubsection{Analysis of the Spreading Factor} The final case study is centered on evaluating whether the proposed model can correctly reproduce the transition from the near-field to the far-field regions of a finite-size RIS. Based on the considered modeling assumptions, the power intensity of the electric field is expected to be constant as a function of the distance in the near-field region, while it is expected to decay with the square of the distance in the far-field region. To validate this trend, we consider an ideal phase-gradient reflector that is illuminated by a normally incident plane wave ($\theta_i=0^\circ$) with a unitary electric field, which is steered towards an anomalous angle of reflection equal to $\theta_r=30^\circ$. Specifically, we compute local averages of the intensity of the electric field over regions of size $10\lambda \times 10\lambda $, and as a function of the observation distances along the direction of the main lobe of the reradiation pattern.

\begin{figure}
\centering
\includegraphics[width=0.4\textwidth]{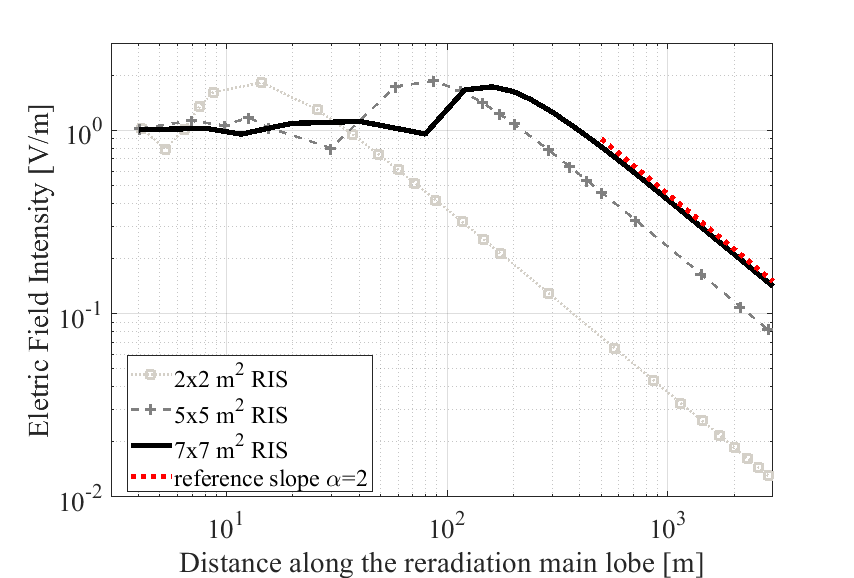}
\caption{Local average of the amplitude of the reradiated electric field by an ideal phase-gradient reflector with $E_i=1$ V/m, $\theta_i=0^\circ$, $\theta_r=30^\circ$. The curves are computed by using (16) as a function of the distance and for different sizes of the RIS at the frequency $f=3$~GHz.}
\label{fig:11} \vspace{-0.5cm}
\end{figure}

\textcolor{black}{The results are shown in Fig.~\ref{fig:11} for different values for the size of the RIS, i.e., $2 \times 2$ m$^2$, $5 \times 5$ m$^2$, and $7 \times 7$ m$^2$, which correspond to $20\lambda \times 20\lambda$, $50\lambda \times 50\lambda$, and $70\lambda \times 70\lambda$, respectively, at the operating frequency of 3 GHz.} \textcolor{black}{From Fig.~\ref{fig:11}, we see that the intensity of the electric field has a behavior that is consistent with the theoretical expectations. Let us consider, for example, an RIS whose size is $7 \times 7$~m$^2$. The average intensity of the electric field remains almost constant up to several tens of meters from the RIS. This is due to the assumption of plane-wave illumination and to the large size of the RIS compared with the distance. If the observation point is located in the radiative near-field region of the RIS, the average intensity of the electric field is characterized by some ripple effects that are determined by the impact of the edge-diffracted waves. If the observation point is located in the Fraunhofer far-field region of the RIS, i.e., at distances greater than approximately 1000~m for the considered case study, the average intensity of the electric field has the typical slope of a spherical wave. The trend is the same for RISs having a smaller size with the only exception that the transition between the two slopes of the curves occur for shorter distances, as expected.} Therefore, we conclude that the proposed approach can model both near-field and far-field propagation regimes.

\section{Conclusion}
\label{sec5}
We have introduced a physically consistent and realistic macroscopic model for evaluating the multi-mode reradiation and diffuse scattering from general engineered reconfigurable  surfaces. The model is based on a hybrid approach, according to which well-established ray-based methods for modeling specular reflection, diffraction, and diffuse scattering, are complemented with the Huygens principle for modeling anomalous reradiated modes. Specifically, ray-based and Huygens-based methods are coupled together through a parametric power balance constraint that ensures the energy conservation between the incident and scattered fields. We have compared two different formulations of the Huygens principle. The first approach is based on the induction theorem and the second approach is based on antenna-array theory. Furthermore, the feasibility and accuracy of both methods have been discussed. In addition, we have implemented the complete macroscopic model and have validated its accuracy against analytical models, full-wave electromagnetic simulations, and experimental measurements available in the literature. Possible generalizations of the proposed model include its use for link-level and system-level performance evaluations in realistic multipath propagation scenarios, as well as the development of a ray-based framework for modeling multi-mode reradiation for different types of reconfigurable surfaces.

\appendices
\section{Generalized method of image currents and derivation of \small{(16)}}
\numberwithin{equation}{section}
By virtue of the induction theorem and the theorem of image currents \cite{ref22, ref23}, the field reradiated by a physical object (a scatterer) that is illuminated by an incident electromagnetic field can be determined by equivalent electric and magnetic surface current densities, which depend on the incident signal and by appropriate image current densities that are in turn determined by the incident signal and the physical properties of the object, i.e., the RIS in our case. In canonical electromagnetic scattering problems, the object is assumed to be either a PEC or a PMC. However, an RIS is a more complex surface, for which it is necessary to consider Maxwell’s equations in the presence of both electric and magnetic currents. For general wave transformations, therefore, the reradiated field  depends on both current densities \cite{ref19}.

The method of current images can be generalized in the presence of an RIS as shown in Fig.~A1. Fig.~A1.a represents the conventional case of a PEC surface element, where the induced electric currents are shorted by the PEC. After applying the method of images and removing the surface, thus, only the magnetic currents are non-zero. On the other hand, Fig.~A1.b depicts the case study of a surface element (e.g., a RIS surface element) that is characterized by a generic surface impedance, and the corresponding modified method of images. In this latter case, electric and magnetic currents are present after removing the surface, and they both depend on the macroscopic coefficient ${\Upgamma}$, which can be interpreted as the local reflection coefficient under the assumption of locally specular reflection. Therefore, the rest of the proof assumes that every single point of the RIS is the source of a wavelet that is locally polarized as a specularly reflected wave and is spatially modulated with the coefficient ${\Upgamma}$. The reradiated wave is obtained, by virtue of Huygens' principle, as a result of the summation of the locally reflected wavelets.
\begin{figure}[!t]
\centering
\includegraphics[width=0.4\textwidth]{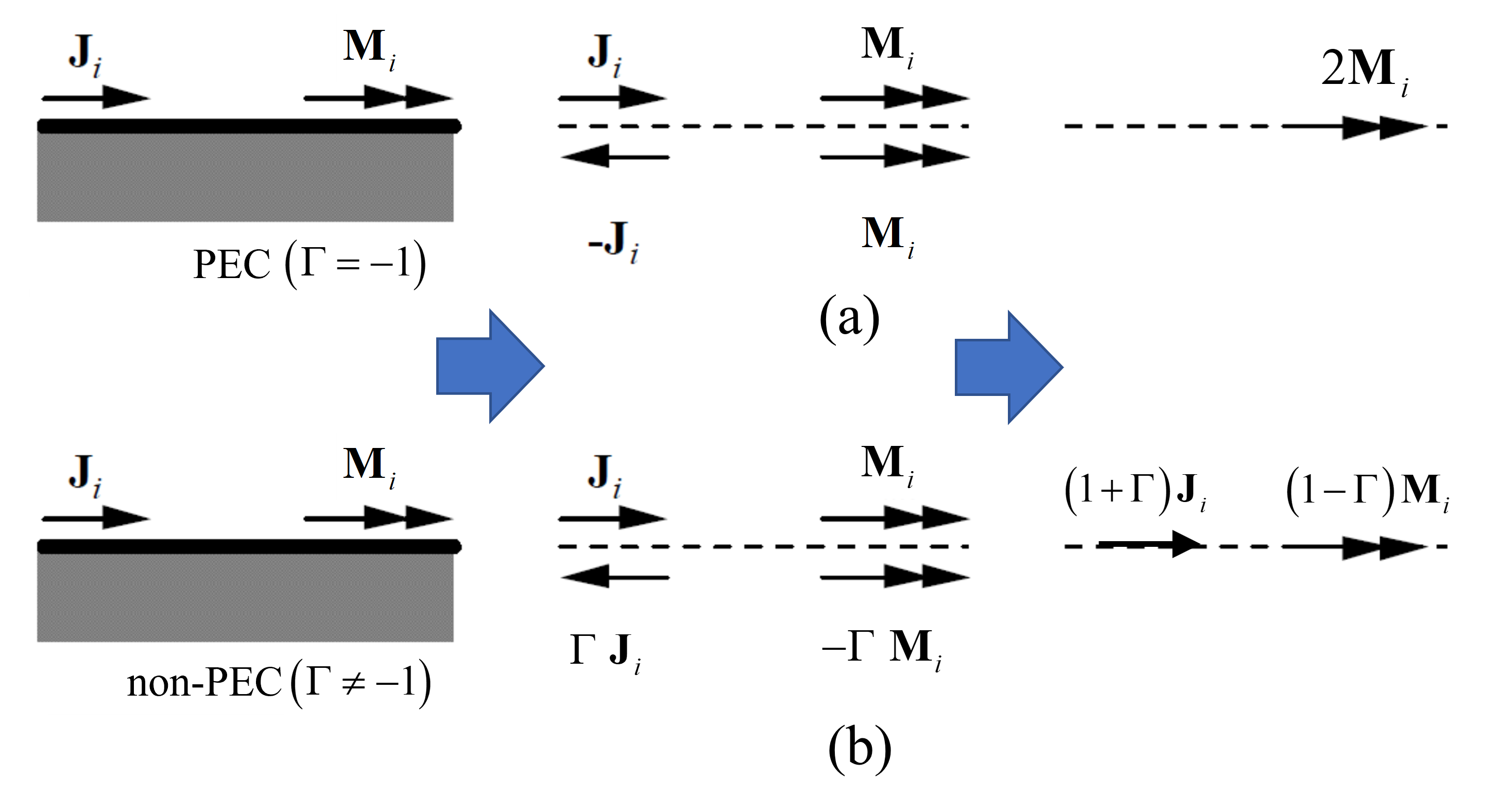}
\caption*{Fig. A1. Conventional (a) and modified (b) method of image currents. The image currents replace the presence of the object (scatterer, i.e., the RIS) for the purpose of calculating the field outside of it.} \vspace{-0.5cm}
\end{figure}

Specifically, according to the induction theorem, the induced surface currents are:
\begin{equation}
\begin{array}{l}
\mathbf{J}_{\mathrm{i}}\left(P'\right)=\mathbf{H}_{i}\left(P'\right)\times \hat{\mathbf{n}}\\
\mathbf{M}_{i}\left(P'\right)=\hat{\mathbf{n}}\times \mathbf{E}_{i}\left(P'\right)
\end{array}
\end{equation}
where $P'=\left(x',y'\right)\in S_{RIS}$ is a generic point on the surface of the RIS, $\hat{\mathbf{n}}$ is the unit normal vector that points outwards (i.e., towards the reflection half-space), and $\boldsymbol{E}_{i}$ and  $\boldsymbol{H}_{i}$ are the incident electric and magnetic fields, respectively. In addition, the image current densities are (see Fig.~A1.b):
\begin{equation}
\begin{array}{l}
\mathbf{J}_{\text{imag}}\left(P'\right)=\Upgamma \left(P'\right)\mathbf{J}_{\mathrm{i}}\left(P'\right)\\
\mathbf{M}_{\text{imag}}\left(P'\right)=-\Upgamma \left(P'\right)\mathbf{M}_{i}\left(P'\right)
\end{array}
\end{equation}
where $\Upgamma \left(P'\right)$ is the macroscopic spatial modulation coefficient in (11) and (12).

Based on (A.1) and (A.2), the field reradiated by the RIS can be calculated by replacing the RIS with the total surface current densities as follows:
\begin{equation}
\begin{array}{l}
\mathbf{J}\left(P'\right)=\left(1+\Upgamma \left(P'\right)\right)\left[\mathbf{H}_{i}\left(P'\right)\times \hat{\mathbf{n}}\right] \\
\mathbf{M}\left(P'\right)=\left(1-\Upgamma \left(P'\right)\right)\left[\hat{\mathbf{n}}\times \mathbf{E}_{i}\left(P'\right)\right]
\end{array}
\end{equation}
and by assuming that the reradiation occurs in the absence of any physical objects, i.e., in free space. The total surface currents in (A.3) assume that the RIS is of infinite extent and the impact of the edges on the surface currents is ignored.

By using the notation in Fig.~3, the electric field that is reradiated by a finite-size RIS in the generic observation point $P=\left(x,y,z\right)$ above the surface can be obtained from Kottler’s formula \cite[Eq.~(18.4.1)]{ref30}:
\begin{equation}
\begin{array}{c}
\mathbf{E}_{\mathrm{m}}\left(P\right)=1/\left(j\omega \varepsilon \right)\int \int _{{S_{RIS}}}k^{2}G\left(x',y'\right)\mathbf{J}\left(x',y'\right)dS\\
\hspace{0.5cm}+1/\left(j\omega \varepsilon \right)\int \int _{{S_{RIS}}}\left(\mathbf{J}\left(x',y'\right)\cdot \nabla '\right)\nabla 'G\left(x',y'\right)dS\\
\hspace{0.5cm}-1/\left(j\omega \varepsilon \right)\int \int _{{S_{RIS}}}j\omega \varepsilon \mathbf{M}\left(x',y'\right)\times \nabla 'G\left(x',y'\right)dS
\end{array}
\end{equation}
where $\nabla '$ denotes the gradient operator with respect to $\left(x',y'\right)$, ``$\times$'' denotes the vector cross product, ``$\cdot $'' denotes the scalar dot product, and $G\left(x',y'\right)$ is the free-space Green's function:
\begin{equation}
G\left(x',y'\right)=e^{-jk\left| \mathbf{r}-\mathbf{r}'\right| }/\left(4\pi \left| \mathbf{r}-\mathbf{r}'\right| \right)=e^{-jkr''}/\left(4\pi r''\right)
\end{equation}
where $\boldsymbol{r}''=\boldsymbol{r}-\boldsymbol{r}'$ and $r''=\left| \boldsymbol{r}''\right| $. It can be observed that the gradient operator $\nabla '$ can be written as follows:\\[2pt]
\centerline {$\nabla ' = \nabla ' \left| \boldsymbol{r}-\boldsymbol{r'}\right| \dfrac{ \partial}{\partial \left| \boldsymbol{r}-\boldsymbol{r'}\right|} = -\dfrac{\boldsymbol{r''}}{\left| \boldsymbol{r''}\right|} \dfrac{\partial}{\partial r''} =  -\hat{\boldsymbol{r}}''\dfrac{\partial}{\partial r''}$} \\[4pt]
when it acts on a function of $r''=\left| \boldsymbol{r}-\boldsymbol{r'}\right| = \sqrt{\boldsymbol{r} \cdot \boldsymbol{r} -2\boldsymbol{r} \cdot \boldsymbol{r'} + \boldsymbol{r'} \cdot \boldsymbol{r'}}$.

Besides, the second and third integrals in (A.4) can be simplified by taking into account that, for typical communication scenarios that encompass the radiative near-field and far-field regions of the RIS, the following approximations hold true:
\begin{equation}
\begin{array}{l}
\nabla ' G\left(x',y'\right) = \left(jk+\dfrac{1}{r''}\right)\dfrac{e^{-jkr''}}{4 \pi r''}\hat{\boldsymbol{r}}'' \approx jkG\left(x',y'\right)\hat{\boldsymbol{r}}'' \\[6pt]
\nabla ' \approx jk\hat{\boldsymbol{r}}''
\end{array}
\end{equation}
where the terms that decay faster than $1/r''$ are neglected.

By using (A.6), the following expression is obtained:
\begin{equation}
\begin{array}{l}
\left(\mathbf{J}\left(x',y'\right)\cdot \nabla '\right)\nabla 'G\left(x',y'\right)\\
\hspace{2cm}\approx -k^{2}G\left(x',y'\right)\left(\mathbf{J}\left(x',y'\right)\cdot \hat{\mathbf{r}}''\right)\hat{\mathbf{r}}''
\end{array}
\end{equation}
Also, the following triple vector product identity holds:
\begin{equation}
\begin{array}{l}
\mathbf{J}\left(x',y'\right)-\left(\mathbf{J}\left(x',y'\right)\cdot \hat{\mathbf{r}}''\right)\hat{\mathbf{r}}'' =  \hat{\mathbf{r}}''\times \left(\mathbf{J}\left(x',y'\right)\times \hat{\mathbf{r}}''\right)
\end{array}
\end{equation}

By inserting (A.6), (A.7) and (A.8) in (A.4), and recalling that $k/\left(\omega  \epsilon \right)=\eta$, the reradiated electric field can be formulated as follows:
\begin{equation}
\begin{array}{l}
\mathbf{E}_{\mathrm{m}}\left(P\right)\approx
-jk\int \int _{{S_{RIS}}}\eta G\left(x',y'\right)\left[\hat{\mathbf{r}}''\times \left(\mathbf{J}\left(x',y'\right)\times \hat{\mathbf{r}}''\right)\right]dS\\
\hspace{1.05cm}-jk\int \int _{{S_{RIS}}} G\left(x',y'\right)\left(\mathbf{M}\left(x',y'\right)\times \hat{\mathbf{r}}''\right)dS
\end{array}
\end{equation}

By substituting (A.3) and (A.5) in (A.9), we eventually obtain (16). The proof follows by noting that the reradiated field in (A.9) is formulated in terms of the surface current densities in (A.3) and it can hence be obtained from the incident wave and the macroscopic coefficient in (11) or (12).

In a similar way, the expression of the reradiated magnetic field can be obtained by applying the approximations (A.6) to the Kottler's formula for the \textbf{H} field \cite[Eq.~(18.4.1)]{ref30}:
\begin{equation}
\begin{array}{l}
\mathbf{H}_{\mathrm{m}}\left(P\right)\approx
-jk\int \int _{{S_{RIS}}}\dfrac{1}{\eta} G\left(x',y'\right)\left[\hat{\mathbf{r}}''\times \left(\mathbf{M}\left(x',y'\right)\times \hat{\mathbf{r}}''\right)\right]dS\\
\hspace{1.05cm}+jk\int \int _{{S_{RIS}}} G\left(x',y'\right)\left(\mathbf{J}\left(x',y'\right)\times \hat{\mathbf{r}}''\right)dS
\end{array}
\end{equation}

Finally, we note that (A.9) and (A.10) can be further simplified in the Fraunhofer far-field region of the RIS, and, in some cases, the reradiated field may be formulated in a closed-form expression. Due to space limitations, the corresponding formulas are not reported but they can be obtained by using the following approximation:
\begin{equation}
G\left(x',y'\right)=\frac{e^{-jk\left| \mathbf{r}-\mathbf{r}'\right| }}{4\pi \left| \mathbf{r}-\mathbf{r}'\right| }\approx \frac{e^{-jk\left| \mathbf{r}\right| }}{4\pi \left| \mathbf{r}\right| }\exp \left(jk\left(\frac{\mathbf{r}}{\left| \mathbf{r}\right| }\cdot \mathbf{r}'\right)\right)
\end{equation}

\bibliographystyle{IEEEtran}

\end{document}